\documentclass[prc,aps,twocolumn]{revtex4}
\usepackage{graphicx}% Include figure files
\usepackage{dcolumn}% Align table columns on decimal point
\usepackage{bm}% bold math

\def\tend{\mathop{\to}}

\def\lim{\mathop{\rm {lim}}}

\begin{document}
\draft \preprint{HEP/123-qed} \widetext
\title{Quantum mechanics and low energy nucleon dynamics}
\author{Renat Kh.Gainutdinov and Aigul A.Mutygullina}
\address{
Department of Physics, Kazan State University, 18 Kremlevskaya St,
Kazan 420008, Russia }
%\email{Renat.Gainutdinov@ksu.ru}
\date{\today}

\begin{abstract}
We discuss the problem of consistency of quantum mechanics as
applied to low energy nucleon dynamics with the symmetries of QCD.
It is shown that the dynamics consistent with these symmetries is
not governed by the Schr{\"o}dinger equation. We present a new way
to formulate the effective theory of nuclear forces as an
inevitable consequence of the basic principles of quantum
mechanics and the symmetries of strong interactions. We show that
being formulated in this way the effective theory of nuclear
forces can be put on the same firm theoretical grounds as the
quantum mechanics of atomic phenomena. In this case the effective
theory allows one to describe with a given accuracy not only
two-nucleon scattering, but also the evolution of nucleon systems,
and places the constraints on the off-shell behavior of the
two-nucleon interaction. In this way we predict the off-shell
behavior of the $S$ wave two-nucleon $T$-matrix at very low
energies when the pionless theory is applicable. Further
extensions and applications of this approach are discussed.
\end{abstract}
 \maketitle
\narrowtext

\section{Introduction}
 Ideas from the foundations of quantum mechanics are
being applied now to many branches of physics. In the quantum
mechanics of particles interacting through the Coulomb potential
one deals with a well-defined interaction Hamiltonian and the
Schr{\"o}dinger equation governing the dynamics of the theory.
This theory is perfectly consistent and provides an excellent
description of atomic phenomena at low energies. It is natural to
expect that low energy nuclear physics can be described in the
same way. However, one has not yet constructed a fundamental
nucleon-nucleon (NN) potential. Nowadays there exist
phenomenological NN potentials which successfully describe
scattering data to high precision, but they do not emerge from QCD
and contain $ad$ $hoc$ form factors. A first attempt to
systematically solve the problem of low energy nucleon dynamics
and construct a bridge to QCD was made by Weinberg \cite{EFT3}. He
suggested to derive a NN potential in time-ordered chiral
perturbation theory (ChPT). However, such a potential is singular
and the Schr{\"o}dinger (Lippmann-Schwinger) equation does not
make sense without regularization and renormalization. This means
that in the effective field theory (EFT) of nuclear forces, which
following the pioneering work of Weinberg has become very popular
in nuclear physics (for a review, see Ref. \cite{rev}), the
Schr{\"o}dinger equation is not valid. On the other hand, the
whole formalism of fields and particles can be considered as an
inevitable consequence of quantum mechanics, Lorentz invariance,
and the cluster decomposition principle \cite{Weinberg97}. Thus in
the nonrelativistic limit QCD must produce low energy nucleon
physics consistent with the basic principles of quantum mechanics.
However, as it follows from the Weinberg analysis, QCD leads
through ChPT to the low energy theory in which the Schr{\"o}dinger
equation is not valid. This means that either there is something
wrong with QCD and ChPT or the Schr{\"o}dinger equation is not the
basic dynamical equation of quantum theory. Meanwhile, in Ref.
\cite{R.Kh.:1999} it has been shown that the Schr{\"o}dinger
equation is not the most general equation consistent with the
current concepts of quantum physics, and a more general equation
of motion has been derived as a consequence of the basic
postulates of the Feynman \cite{Feynman:1948} and canonical
approaches to quantum theory. Being equivalent to the
Schr{\"o}dinger equation in the case of instantaneous
interactions, this generalized dynamical equation permits the
generalization to the case where the dynamics of a system is
generated by a nonlocal-in-time interaction. The generalized
quantum dynamics (GQD) developed in this way has proved an useful
tool for solving various problems in quantum theory
\cite{PRC:2002,PLA:2002}.

The formalism of the GQD allows one to consider the problem of
consistency of quantum mechanics with the low energy predictions
of QCD from a new  point of view. From this viewpoint, in
investigating the consequences of ChPT we must not restrict
ourselves to the assumption that the $NN$ interaction can be
parameterized by a $NN$ potential, and low energy nucleon dynamics
is governed by the Schr{\"o}dinger equation. This dynamics may be
governed by the generalized dynamical equation with a
nonlocal-in-time interaction operator when this equation is not
equivalent to the Schr{\"o}dinger equation, and hence the above
divergence problems may be the cost of trying to describe low
energy nucleon dynamics in terms of Hamiltonian formalism while
this dynamics is really non-Hamiltonian. In the present paper, by
using the example of the ${}^1S_0$ channel, we will show that from
the analysis of time-ordered diagrams for the $T$-matrix in ChPT
it really follows that  nucleon dynamics at very low energies is
governed by the generalized dynamical  equation with a
nonlocal-in-time interaction operator.
 The GQD provides a new way to formulate the effective
theory of nuclear forces as an inevitable consequence of the basic
principles of quantum mechanics and the symmetries of QCD,
 and, as it will be shown,
in this way one can construct the theory that is perfectly
consistent and free from ultraviolet (UV) divergences.

The aim of the present paper is twofold. From the one hand, we
intend to show that the low energy predictions of QCD are
consistent with the basic principles of quantum mechanics, but in
this case a new insight into these principles provided by the
formalism of the GQD is needed. On the other hand, it is our
intention to demonstrate that this insight into the basic
principles of quantum mechanics opens new possibilities for
describing low energy nucleon dynamics. In Sec.II we briefly
consider the main features of the formalism of the GQD developed
by one of the authors (R.G.) in Ref. \cite{R.Kh.:1999}. We mainly
focus on the physical meaning of the generalized dynamical
equation which is a direct consequence of the principle of the
superposition of the probability amplitudes and the requirement
that the evolution operator be unitary. This equation plays a key
role in the present work.

In Sec.III we consider the pionless theory of nuclear forces. By
focusing the attention on the ${}^1S_0$  channel, we show that the
requirement that the two-nucleon $(2N)$ $T$-matrix satisfy the
generalized dynamical equation and be consistent with the
symmetries of QCD completely determines its form. This means that
the pionless effective theory can be formulated as an inevitable
consequence of the symmetries of QCD and the basic principles of
quantum mechanics, if this principles are entered into the theory
via the generalized dynamical equation. In Sec.IV we show that,
starting from the expression for the $2N$  $T$-matrix obtained in
this way one can organize calculations of observables in the
spirit of the effective theory, i.e., by expanding amplitudes in
powers of $Q/\Lambda$ where $Q$ being some low energy scale, and
$\Lambda$ is the scale at which the theory is expected to break
down. This makes it possible to calculate amplitudes up to some
order in $Q/\Lambda$ keeping the terms in this expansion up to the
corresponding order. In this way we reproduce all results for the
$NN$ scattering amplitudes obtained in the standard EFT of nuclear
forces. It is important that these results are reproduced starting
with a well defined $2N$ $T$-matrix without resorting to the
regularization and renormalization procedures. This shows that
really the ChPT does not lead, in the nonrelativistic limit, to a
low energy theory with UV divergences, i.e., low energy
predictions of QCD are consistent with the principles of quantum
mechanics.

In Sec.V we show that the use of the generalized dynamical
equation allows one to formulate the effective theory as a
perfectly consistent theory free from UV divergences. Being
formulated in this way, the effective theory keeps all
 advantages
of the traditional nuclear physics approach. At the same time, its
advantage over the traditional approach is that it allows one to
find constraints on the off-shell behavior of the $2N$ $T$-matrix
that are placed by the symmetries of QCD. In contrast, the
realistic potentials that has been developed within the
traditional approach cannot guarantee a reliable off-shell
extrapolation of the $2N$ $T$-matrix since they are all
constrained by the $2N$ phase shift analysis.

In Sec.VI we show that, in despite of the wide-spread opinion,
from the first principles of quantum mechanics it follows that in
general renormalization is not necessary to separate the low
energy physics that is described by the effective theory of
nuclear forces from the underlying high energy physics. An
important feature of the generalized dynamical equation is that it
provides such a separation without renormalization. The only thing
that is needed for this is that at low energies nucleons emerge as
the only effective degrees of freedom.

In Sec.VII the proposed formalism is considered from the point of
view of the Weinberg program for physics of the two-nucleon
systems. We show that our formulation of the effective theory of
nuclear forces can be considered as a new way of the realization
of the Weinberg program. This way leads to a perfectly consistent
effective theory that allows one not only to calculate the $2N$
scattering amplitudes but $2N$ $T$-matrix and hence the evolution
and Green operators. Moreover, the effective theory developed in
this way allows one to determine the off-shell behavior of the
$2N$ $T$-matrix as an inevitable consequence of the first
principles of quantum mechanics and the symmetries of QCD. This is
very important, because, as is well known, the off-shell behavior
of the $2N$ $T$-matrix may play a crucial role in solving the
many-nucleon problem and is an important factor in calculating
in-medium observables \cite{Fuchs} and in microscopic nuclear
structure calculations. This results, for example, in the fact
that the predictions by the Bonn potential for nuclear structure
problems differ in a characteristic way from the ones obtained
with local realistic potentials \cite{Machleidt}.  The off-shell
ambiguities of realistic $NN$ potentials are argued to be one of
the main causes of many problems in describing three-nucleon
systems. One of such problems is the $A_y$ puzzle which refers to
the inability to explain the nucleon vector analyzing power $A_y$
in elastic nucleon-deuteron ($Nd$) scattering at low energy. As
has been shown in Ref. \cite{Huber}, it is not possible with
reasonable changes in realistic $NN$ potentials to increase the
$Nd$ $A_y$ and at the same time to keep the $2N$ observables
unchanged. The same situation also takes place in the case of
chiral potentials \cite{Entem}. This means that the introduction
of three-nucleon forces (3NF) is needed for resolving the problem.
However, as it turned out, conventional $3N$F's change the
predictions for $Nd$ $A_y$ only slightly and do not improve them
[12-14]. This motivated new types of $3N$F's [15-17]. On the other
hand, reliable $3N$ calculations and even testing for the presence
of $3N$ forces require to constrain the off-shell properties of
the $2N$. Moreover, these properties play a crucial role in a new
$3N$F proposed by Canton and Schadow \cite{Canton}. These aspects
of the three-nucleon problem are discussed in Sec.VI. We show that
the fact that the GQD allows one to predict the off-shell behavior
of the $2N$ $T$-matrix, by using the symmetry constrains placed by
a QCD, may open new possibilities for solving three-body problem
in nuclear physics.

\section{Generalized quantum dynamics}

The basic concept of the canonical formalism of quantum theory is
that it can be formulated in terms of vectors of a Hilbert space
and operators acting on this space. In this formalism the
postulates that establish the connection between the vectors and
operators and states of a quantum system and observables are used
together with the dynamical postulate according to which the time
evolution of a quantum system is governed by the Schr{\"o}dinger
equation. In the Feynman formalism quantum theory is formulated in
terms of probability amplitudes without resorting to vectors and
operators acting on a Hilbert space. Feynman's theory starts with
an analysis of the phenomenon of quantum interference. The results
of this analysis which leads directly to the concept of
 the superposition of probability amplitudes are summarized by
the following postulate \cite{Feynman:1948}:

 The probability of an event is the absolute square of a
complex number called the probability amplitude. The joint
probability amplitude of a time-ordered sequence of events is the
product of separate probability amplitudes of each of these
events. The probability amplitude of an event which can happen in
several different ways is a sum of the probability amplitudes for
each of these ways.

According to this assumption, the probability amplitude of an
event which can happen in several different ways is a sum of
contributions from each alternative way.
%Dividing these
%alternatives in different classes, one can then analyze such a
%probability amplitudes in different ways.
The Feynman formulation is based on the assumption that the
history of a quantum system can be represented by some path in
space-time, and hence the probability amplitude of any event is a
sum of the probability amplitudes that a particle has a completely
specified path in space-time. The contribution from a single path
is postulated to be an exponential whose (imaginary) phase is the
classical action (in units of $\hbar$) for the path in question.
This assumption is not as fundamental as the above principle of
the superposition of probability amplitudes which follows directly
from the analyzes of the phenomenon of quantum interference. This
fact is emphasized, for example, in Feynman's book \cite{Feyn},
where a minimal set of physical principles which must be satisfied
in any theory of fields and particles is analyzed. Feynman
includes in this set neither the second postulate of his formalism
nor the Schr{\"o}dinger equation: The only  quantum mechanical
principle included in this set is the principle of the
superposition of probability amplitudes. In Ref. \cite{R.Kh.:1999}
it has been shown that, instead of processes associated with a
completely specified path in space-time, one can use processes
associated with completely specified instants of the beginning and
end of interaction in a quantum system as alternative ways in
which any event can happen. As it turned out, employing this class
of alternatives allows one to derive a dynamical equation from the
principle of the superposition without making supplementary
assumptions like the second postulate of Feynman's theory.
 By using this class of
alternatives and the superposition principle,
$\langle\psi|U(t,t_0)|\varphi\rangle$, being the probability
amplitude of finding a quantum system in the state $|\psi\rangle$
in a measurement at time $t$,
 if at time $t_0$ it was in the state
 $|\varphi\rangle$, can  be represented in the form \cite{R.Kh.:1999}
\begin{eqnarray}
\langle\psi| U(t,t_0)|\varphi\rangle =
\langle\psi|\varphi\rangle %\nonumber\\
+ \int_{t_0}^t dt_2 \int_{t_0}^{t_2} dt_1 \langle\psi|\tilde
S(t_2,t_1)|\varphi\rangle. \label{ev}
\end{eqnarray}
Here $\langle\psi|\tilde S(t_2,t_1)|\varphi\rangle$ is the
probability amplitude that, if at time $t_1$ the system was in the
state $|\varphi\rangle,$ then the interaction in the system will
begin at time $t_1$ and  end at time $t_2$ and at this time the
system will be in the state $|\psi\rangle.$ By using the operator
formalism, one can represent amplitudes $\langle\psi|
U(t,t_0)|\varphi\rangle$ by the matrix elements of the unitary
evolution operator $U(t,t_0)$ in the interaction picture. The
operator $\tilde S(t_2,t_1)$ represents the contribution to the
evolution operator from the process in which the interaction in
the system begins at time $t_1$ and ends at time $t_2$.  As has
been shown in Ref. \cite{R.Kh.:1999}, for the evolution operator
(\ref{ev}) to be unitary for any $t$ and $t_0$ the operator
$\tilde S(t_2,t_1)$ must satisfy the equation
\begin{eqnarray}
&&(t_2-t_1) \tilde S(t_2,t_1) \nonumber\\
&=& \int^{t_2}_{t_1} dt_4 \int^{t_4}_{t_1}dt_3 (t_4-t_3) \tilde
S(t_2,t_4) \tilde S(t_3,t_1). \label{main}
\end{eqnarray}
A remarkable feature of this equation is that it works as a
recurrence relation and allows one to obtain the operators $\tilde
S(t_2,t_1)$ for any $t_1$ and $t_2$, if $\tilde S(t'_2, t'_1)$
corresponding to infinitesimal duration times $\tau = t'_2 -t'_1$
of interaction are known. It is natural to assume that most of the
contribution to the evolution operator in the limit $t_2\to t_1$
comes from the processes associated with the fundamental
interaction in the system under study. Denoting this contribution
by $H_{int}(t_2,t_1)$ we can write
\begin{equation}
\tilde{S}(t_2,t_1) \tend\limits_{t_2\rightarrow t_1}
H_{int}(t_2,t_1) + o(\tau^{\epsilon}),\label{fund}
\end{equation}
where $\tau=t_2-t_1$. The parameter $\varepsilon$ is determined by
demanding that $H_{int}(t_2,t_1)$  called the generalized
interaction operator must be so close to the solution of Eq.
(\ref{main}) in the limit $t_2\tend t_1$ that this equation has a
unique solution having the behavior (\ref{fund}) near the point
$t_2=t_1$. If $H_{int}(t_2,t_1)$ is specified, Eq. (\ref{main})
allows one to find the operator $\tilde S(t_2,t_1)$, and hence the
evolution operator. Thus Eq. (\ref{main}) which is a direct
consequence of the principle of the superposition can be regarded
as an equation of motion for states of a quantum system.  This
equation allows one to construct the evolution operator by using
the contributions from fundamental processes as building blocks.
In the case of Hamiltonian dynamics the fundamental interaction is
instantaneous. The generalized interaction operator describing
such an interaction is of the form
\begin{equation}
 H_{int}(t_2,t_1) = - 2i \delta(t_2-t_1)
 H_{I}(t_1) \label{loc}
\end{equation}
(the delta function $\delta(t_2-t_1)$ emphasizes that the
interaction is instantaneous). In this case Eq. (\ref{main}) is
equivalent to the Schr{\"o}dinger equation \cite{R.Kh.:1999} and
the operator $H_I(t)$ is an interaction Hamiltonian (in the
interaction picture). At the same time, Eq. (\ref{main}) permits
the generalization to the case where the fundamental interaction
in
 a quantum system is nonlocal in time, and hence
the dynamics is non-Hamiltonian \cite{R.Kh.:1999}.

 By
using Eq. (\ref{ev}), for $U(t,t_0)$, we can write
\begin{eqnarray}
U(t,t_0)&=&  {\bf 1} + \frac{i}{2\pi}
\int^\infty_{-\infty} dx\exp[-i(z-H_0)t]\label{evo}\\
&\times &(z-H_0)^{-1}T(z)(z-H_0)^{-1} \exp[i(z-H_0)t_0],\nonumber
\end{eqnarray}
 where $z=x+iy$, $y>0$, and $H_0$ is the free Hamiltonian.
 The
 operator $T(z)$ is defined by
\begin{eqnarray}
T(z) = i \int_{0}^{\infty} d\tau \exp(iz\tau)\tilde
T(\tau),\label{tz}
\end{eqnarray}
where $\tilde T(\tau)=\exp(-iH_0t_2)\tilde
S(t_2,t_1)\exp(iH_0t_1)$, and $\tau=t_2-t_1$.
 In terms of the
T-matrix defined by Eq. (\ref{tz}) the equation of motion
(\ref{main}) can be rewritten in the form \cite{R.Kh.:1999}
\begin{equation}
\frac{d \langle \psi_2|T(z)|\psi_1\rangle}{dz} = -  \sum
\limits_{n} \frac{\langle \psi_2|T(z)|n\rangle\langle
n|T(z)|\psi_1\rangle}{(z-E_n)^2},\label{difer}
\end{equation}
where $n$ stands for the entire set of discrete and continuous
variables that characterize the system in full, and $|n\rangle$
are the eigenvectors of $H_0$. As it follows from Eq. (\ref{loc}),
the boundary condition on this equation is of the form
\begin{eqnarray}
\langle\psi_2|T(z)|\psi_1\rangle \tend \limits_{|z|\tend\infty} \langle\psi_2|B(z)|\psi_1\rangle
+o\left(|z|^{-\beta}\right)\nonumber\\
=\langle\psi_2|B(z)|\psi_1\rangle
+O\left\{h(z)\right\},\label{T(z)to}\\
B(z) = i \int_{0}^{\infty} d\tau \exp(iz\tau) H_{int}^{(s)}(\tau)
\label{B(z)-Hint}
\end{eqnarray}
 where
$$H^{(s)}_{int}(t_2-t_1) = \exp(-iH_0t_2) H_{int}(t_2,t_1)
\exp(iH_0t_1)$$ is the interaction operator in the Schr{\"o}dinger
picture, and $h(z)$ is an arbitrary function satisfying the
condition $h(z)=o\left(|z|^{-\beta}\right)$, $|z|\to\infty$, with
$\beta=\varepsilon+1$. In the case of the Hamiltonian dynamics
when $H_{int}^{(s)}(\tau)=-2i\delta(\tau)H_I$, with $H_I$ being
the interaction Hamiltonian in the Schr{\"o}dinger picture,
 $B(z)=H_I$, and
the boundary condition (\ref{T(z)to}) takes the form
\begin{eqnarray}
T(z) \tend \limits_{|z|\tend\infty} H_I. \label{B(z)-H_I}
\end{eqnarray}
Equation (\ref{difer}) with this boundary condition is equivalent
to the Lippmann-Schwinger (LS) equation with the interaction
Hamiltonian $H_I$.
 By definition, the operator $B(z)$
represents the contribution which $H_{int}^{(s)}(\tau)$ gives to
the operator $T(z)$, and is the interaction operator in the energy
representation. This operator must be so close to the relevant
solution of Eq. (\ref{difer}) in the limit $|z|\to\infty$ that
this differential equation has a unique solution having the
asymptotic behavior (\ref{T(z)to}). For this the operator $B(z)$
must satisfy the condition
\begin{eqnarray}
\frac{d \langle \psi_2|B(z)|\psi_1\rangle}{dz} &=& - \sum
\limits_{n} \frac{\langle \psi_2|B(z)|n\rangle\langle n |B(z)|\psi_1\rangle}{(z-E_n)^2}\nonumber\\
&+&o(|z|^{-\beta-1}), \qquad |z|\to\infty. \label{diferB}
\end{eqnarray}

From Eq. (\ref{evo}) it follows that the evolution operator in the
Schr{\"o}dinger picture can be represented in the form
\begin{eqnarray}
U_s(t,0)=\frac{i}{2\pi}\int\limits_{-\infty}^{\infty}dx\exp(-izt)
G(z),\label{U-G}
\end{eqnarray}
where $z=x+iy$, $y>0$, and
\begin{eqnarray}
 G(z)=G_0(z)+G_0(z)T(z)G_0(z)\label{G-T}
\end{eqnarray}
with $G_0(z)=\frac{1}{z-H_0}.$ Being equivalent to representation
(\ref{ev}), Eqs. (\ref{U-G}) and (\ref{G-T}) express the principle
of the superposition of probability amplitudes. In the ordinary
quantum mechanics the similar equation plays an important role and
establishes the connection between the evolution operator and the
Green operator $G(z)$ which is defined by
\begin{eqnarray}
 G(z)=\frac{1}{z-H},\label{G}
\end{eqnarray}
with $H$ being the total Hamiltonian. Such a form of the Green
operator follows from the fact that in the Hamiltonian formalism
the evolution operator satisfies the Schr{\"o}dinger equation. In
the canonical formalism the $T$-matrix is defined by Eq.
(\ref{G-T}) starting with the Green operator of the form
(\ref{G}). In the formalism of the GQD the $T$-matrix plays a more
fundamental role. It is defined by Eq. (\ref{tz}), and in this
case the starting point is representation (\ref{ev}) with $\tilde
S(t_2,t_1)$ being the contribution to the evolution operator from
the processes in which the interaction begins at time $t_1$ and
ends at time $t_2$. In this case the operator $G(z)$ itself is
defined by Eq. (\ref{G-T}) via the $T$-matrix. This is a more
general definition of the Green operator, since representation
(\ref{ev}) is a consequence of the principle of the superposition
of probability amplitudes and must be valid in any case while the
evolution operator can be represented in the form (\ref{U-G}) with
the operator $G(z)$ given by Eq. (\ref{G}) only in the case where
the interaction in a system is instantaneous.

As has been shown in Ref. \cite{PRC:2002}, there is a one-to-one
correspondence between the character of the dynamics and the
large-momentum behavior of the $T$-matrix. If this behavior
satisfies the requirements of ordinary quantum mechanics, then the
interaction in a system is instantaneous and the dynamics is
Hamiltonian. In the case where this behavior is "bad", i.e., does
not meet the requirements of the ordinary quantum mechanics, the
interaction generating the dynamics of the system must necessarily
be nonlocal-in-time. Let us now illustrate this point by using the
model developed in Refs. \cite{R.Kh.:1999,PRC:2002} as a test
model demonstrating the possibility of the extension of quantum
dynamics provided by the GQD. This model describes the evolution
of the system of two identical nonrelativistic particles whose
interaction is separable, and hence the interaction operator has
the form
\begin{eqnarray}
 <{\bf p}_2| H_{int}^{(s)}(\tau)|{\bf p}_1>
 = \varphi^*({\bf p}_2)\varphi({\bf p}_1)f(\tau),\label{ftau}
\end{eqnarray}
where ${\bf p}$ is the relative momentum of the particles, and
$f(\tau)$ is some function of the duration time $\tau$ of the
interaction in the system. It is assumed that in the limit $|{\bf
p}|\to\infty$ the form factor $\varphi({\bf p})$ behave as
\begin{equation}
\varphi({\bf p}) \sim |{\bf p}|^{-\alpha}, \quad {(|{\bf p}| \tend
\infty).} \label{form}
\end{equation}
In this case the general solution of Eq. (\ref{difer}) is
\cite{PRC:2002}
\begin{eqnarray}
\langle{\bf p}_2|T(z)|{\bf p}_1\rangle =\frac{ \varphi^* ({\bf
p}_2)\varphi({\bf p}_1)}{g_a^{-1} +(z-a) \int
\frac{d^3k}{(2\pi)^3} \frac {|\varphi({\bf
k})|^2}{(z-E_k)(a-E_k)}},\label{sol}
\end{eqnarray}
where $g_a=t(a)$, and $a\in(-\infty,0]$. In the case
$\alpha>\frac{1}{2}$, the amplitude $\langle{\bf p_2}|T(z)|{\bf
p_1}\rangle$ given by Eq. (\ref{sol}), tends to $\lambda
 \varphi^*({\bf p}_2)\varphi({\bf p}_1)$, where
$$\lambda=\left(g_a^{-1}+\int\frac{d^3k}{(2\pi)^3}
\frac{|\varphi({\bf k})|^2}{a-E_k}\right)^{-1}.$$ From Eqs.
(\ref{T(z)to}) and (\ref{B(z)-Hint}) it follows that in this case
the interaction operator $H_{int}^{(s)}(\tau)$ should be of the
form
$$<{\bf p}_2| H_{int}^{(s)}(\tau)|{\bf p}_1>
 = -2i\delta(\tau)\varphi^*({\bf p}_2)\varphi({\bf p}_1),
$$
and hence the dynamics of the system is Hamiltonian and is
governed by the Schr{\"o}dinger equation with the potential
$\langle{\bf p_2}|V|{\bf p_1}\rangle=\lambda
 \varphi^*({\bf p}_2)\varphi({\bf p}_1)$.

 In the case $\alpha<\frac{1}{2}$, the $T$-matrix (\ref{sol})
 tends to zero as $|z|\tend\infty$
 \begin{eqnarray}
\langle{\bf p_2}|T(z)|{\bf p_1}\rangle \tend \limits_{|z| \tend
\infty} \varphi^* ({\bf p}_2)\varphi({\bf p}_1)\nonumber\\
\times\left( b_1 (-z)^{\alpha-\frac{1}{2}}+ b_2 (-z)^{2
\alpha-1}\right) + o(|z|^{2 \alpha-1}),\label{tzero}
\end{eqnarray}
where $b_1 =- 4\pi \cos(\alpha \pi) m^{\alpha-\frac{3}{2}}$ and
$b_2= b_1 |a|^{\frac{1}{2}- \alpha} -b_1^2(\tilde M(a)+g_a^{-1})$
with
$$
\tilde M(a) = \int\frac{d^3k}{(2\pi)^3} \frac {|\varphi({\bf
k})|^2-
 |{\bf {k}}|^{-2\alpha}}
{a-E_k} .
$$
In this case the dynamics is non-Hamiltonian and, as it follows
from Eqs. (\ref{T(z)to}) and (\ref{B(z)-Hint}), is generated by
the nonlocal-in-time interaction operator
\begin{eqnarray} <{\bf p}_2| {H}^{(s)}_{int}(\tau)|{\bf p}_1>
& = &\varphi^* ({\bf p}_2)\varphi ({\bf p}_1)\nonumber\\
&\times&\left( a_1\tau^{-\alpha-\frac{1}{2}} + a_2 \tau^{-2
\alpha}\right), \label{less}
\end{eqnarray}
where $a_1= 4\pi i\cos(\alpha \pi) m^{\alpha-\frac{3}{2}} \Gamma
^{-1}(\frac{1}{2}-\alpha) \exp[i(-\frac{\alpha}{2}+ \frac{1}{4})
\pi]$, and $a_2$ is a free parameter of the theory. The solution
of Eq. (\ref{difer}) with the interaction operator (\ref{less}) is
of the form
\begin{eqnarray}
%&&
\langle{\bf p}_2| T(z)|{\bf p}_1\rangle =\frac{b_1^2\varphi^*
({\bf p}_2)\varphi ({\bf
p}_1)}{-b_2+b_1(-z)^{\frac{1}{2}-\alpha}-\tilde
M(z)b_1^2}.\label{nz}
\end{eqnarray}

\section{The pionless theory of nuclear forces}

The Weinberg program for low energy nucleon physics employs the
analysis of time-ordered diagrams for the $2N$ $T$-matrix in ChPT
to derive a $NN$ potential and then to use it in the LS equation
for constructing the full $NN$ $T$-matrix. Obviously the starting
point for this program is the assumption that in the
nonrelativistic limit ChPT leads to low energy nucleon dynamics
which is Hamiltonian and is governed by the Schr{\"o}dinger
equation. However, the fact that the chiral potentials constructed
in this way are singular and lead to UV divergences means that
this assumption has not corroborated. At the same time, the GQD
allows one to analyze the  predictions of ChPT without making $a$
$priori$ assumptions about the character of low energy nucleon
dynamics: This character should results from the analysis. Let us
consider, for  example, the low energy predictions of ChPT for the
$2N$ system in the ${}^1S_0$ channel. At very low energy, even the
pion field can be integrated out, and the diagrams of the ChPT
take the form of the diagrams being produced by the effective
Lagrangian containing only contact interactions among nucleons and
derivatives thereof \cite{Kaplan2}. From the analysis of
time-ordered diagrams of this theory it follows that the $2N$
$T$-matrix in the ${}^1S_0$ channel must be of the form
\begin{equation}
\langle {\bf p_2}|T(z)|{\bf p_1}\rangle = \sum
\limits_{n,m=0}^\infty p_2^{2n}p_1^{2m}t_{nm}(z), \label{pp}
\end{equation}
where ${\bf p}_i$ is relative momentum of nucleons, and the terms
$t_{nm}(z)$ are of order $|t_{nm}(z)|\sim
O\left\{\Lambda^{-2(n+m)}\right\}$, with the expansion parameter
$\Lambda$ being set by the pion mass. Obviously, chiral symmetry
play no role in this case. However, for our analysis it is
important that at extreme low energies ChPT gives rise to this
result.

From Eq. ({\ref{pp}) it follows that at leading order the $2N$
$T$-matrix is of the form
\begin{equation}
\langle{\bf p}_2|T^{(0)}(z)|{\bf p}_1\rangle=t_{00}(z).
\label{t00}
\end{equation}
On the other hand, this $T$-matrix must satisfy the generalized
equation of motion, and this requirement determines its form up to
one arbitrary parameter \cite{PRC:2002}
\begin{equation}
\langle{\bf p}_2|T^{(0)}(z)|{\bf
p}_1\rangle=\left(C_0^{-1}-\frac{m\sqrt{-zm}}{4\pi}\right)^{-1},
\label{t0}
\end{equation}
where $m$ being nucleon mass. Note that the standard EFT approach
yields the same expression for the leading order $T$-matrix
\cite{Kaplan,vKolck}. It is remarkable that the $T$-matrix
(\ref{t0}), which in the EFT approach is obtained by performing
regularization and renormalization of the solution of the
Schr{\"o}dinger (LS) equation, is not a solution of this equation.
At the same time, it is a solution of Eq. (\ref{difer})
corresponding to the interaction operator \cite{PRC:2002}
\begin{equation}
\langle{\bf p}_2| H_{int}^{(s)}(\tau)|{\bf p}_1\rangle =4\pi i
m^{-3/2}\left(\tau^{-1/2}+i\gamma\right), \label{hint}
\end{equation}
where $\gamma=4\pi\exp(-i\pi/4)m^{-3/2}C_0^{-1}$. This operator is
nonlocal in time and this is the only reason why in this case the
generalized dynamical equation cannot be reduced to the LS
equation, and hence the T-matrix is not its solution.

Let us now consider the full $2N$ $T$-matrix including all the
terms in Eq. (\ref{pp}). From Eqs. (\ref{difer}) and (\ref{pp}) it
follows that the functions $t_{nm}(z)$ satisfy the equations
\begin{equation}
\frac{d t_{nm}(z)}{dz} = - \int\frac{d^3k}{(2\pi)^3}
\sum\limits_{n',m'=0}^\infty
\frac{t_{nn'}(z)t_{m'm}(z)k^{2(n'+m')}}{(z-E_k)^2}.\label{d1}
\end{equation}
The solution of this equation is of the form
\begin{equation}
t_{nm}(z) = c'_{2n}c_{2m}t(z), \label{cc}
\end{equation}
where $t(z)$ satisfies the equation
\begin{equation}
\frac{d t(z)}{dz} = - t^2(z)\int\frac{d^3k}{(2\pi)^3}
\frac{\psi_{2}(k/\Lambda)\psi_{1}(k/\Lambda)}{(z-E_k)^2},\label{d2}
\end{equation}
with
\begin{eqnarray}
\psi_1(k/\Lambda)=1+\sum \limits_{n=1}^\infty
c_{2n}k^{2n},\nonumber\\
\psi_2(k/\Lambda)=1+\sum \limits_{n=1}^\infty
c'_{2n}k^{2n}.\nonumber
\end{eqnarray}
The form of the functions $\psi_1(k/\Lambda)$ and
$\psi_2(k/\Lambda)$ manifests the fact that the constants $c_{2n}$
and $c'_{2n}$ are of order $O\{\Lambda^{-2n}\}$, and hence the
relevant domain of the definition of these functions corresponds
to the scale $\Lambda$. As has been shown in Ref.
\cite{R.Kh.:1999}, for the T-matrix being a solution of Eq.
(\ref{d1}) to be unitary, it must satisfy the condition
\begin{eqnarray}
\langle{\bf p}_2| T^+(z)|{\bf p}_1\rangle=\langle{\bf p}_2|
T(z)|{\bf p}_1\rangle,\quad z\in (-\infty,0).\nonumber
\end{eqnarray}
From this and Eq. (\ref{d2}) it follows that
\begin{eqnarray}
\psi_2(k/\Lambda)=\psi^*_1(k/\Lambda),\qquad
c^*_{2n}=c'_{2n}.\nonumber
\end{eqnarray}
Now we can rewrite Eq. (\ref{d2}) in the form
\begin{eqnarray}
\frac{d \tilde t(z)}{dz} = \int\frac{d^3k}{(2\pi)^3}
\frac{1}{(z-E_k)^2}+\int\frac{d^3k}{(2\pi)^3}
\frac{|\psi(k/\Lambda)|^2-1}{(z-E_k)^2},\nonumber
\end{eqnarray}
where $\tilde t(z)=t^{-1}(z)$ and
\begin{eqnarray}
\psi(k/\Lambda)\equiv\psi_1(k/\Lambda)=\sum \limits_{n=0}^\infty
c_{2n}k^{2n},\label{psi=psi1}
\end{eqnarray}
 with $c_0=1$. The solution of this
equation with the boundary condition $\tilde t(z=0)=\tilde t_0$ is
\begin{eqnarray}
 \tilde t(z) &=& \tilde t_0
 -zm^2\int\frac{d^3k}{(2\pi)^3} \frac{1}{(zm-k^2)k^2}-M(z),
\nonumber
\end{eqnarray}
where
$$M(z) =zm^2\int \frac{d^3 k}{(2\pi)^3} \frac{F_1(k/\Lambda)}
{(zm-k^2)k^2},$$ with $F_1(k/\Lambda)=|\psi(k/\Lambda)|^2-1$.
Hence
\begin{eqnarray}
 t(z) =
 \left[C_0^{-1}-\frac{m}{4\pi}\sqrt{-zm}-M(z)\right]^{-1},\nonumber
% \label{solu}
\end{eqnarray}
where $C_0=\tilde t_0^{-1}$. Thus from the requirement that the
$2N$ T-matrix having the structure (\ref{pp}) satisfy the
generalized dynamical equation it follows that this T-matrix
should be of the form
\begin{equation}
\langle{\bf p}_2| T(z)|{\bf p}_1\rangle
=\frac{\psi^*(p_2/\Lambda)\psi(p_1/\Lambda)}{C_0^{-1}-\frac{m}{4\pi}\sqrt{-zm}-M(z)}.\label{T-}
\end{equation}

\section{The expansion of the effective theory.}

As we have shown, the requirement that low energy nucleon dynamics
be governed by the generalized dynamical equation places
additional constraints on the form of the $2N$ $T$-matrix. This
requirement together with the requirement that in the ${}^1S_0$
channel the $2N$ $T$-matrix have the structure (\ref{pp}) yield
the expressions (\ref{T-}) for this $T$-matrix which can be used
for constructing the scattering amplitude and the evolution
operator. The only problem is that we do not know the values of
the parameters $c_{2n}$. One may hope that in future it will be
possible to derive them from QCD. Another way is to obtain these
parameters from low energy experiment. Of course, in this way one
has to restrict oneself to a few parameters that should be fitted
to experiment. In other words, one has to use  Eq. (\ref{T-}) in
the spirit of the effective theory of nuclear forces, keeping only
terms in the theory up to some order in $Q/\Lambda$ in
calculations of observables up to the corresponding order.

In the EFT approach the scattering amplitude, which is given by
the equation
\begin{equation}
A(p)=\frac{4\pi}{m}\frac{1}{p\cot\delta-ip}\label{Ap}
\end{equation}
(here and below we focus on the ${}^1S_0$ partial wave), is
described in terms of an expansion in powers of momentum over the
characteristic short-distance scale $\Lambda$. It is well known
that the quantity $p\cot\delta$ has a nice momentum expansion for
$p<<\Lambda$ known as the The Effective Range Expansion (ERE):
\begin{equation}
p\cot\delta=-\frac{1}{a}+\frac{1}{2}\sum \limits_{n=0}^{\infty}r_n
p^{2(n+1)}.\label{cot}
\end{equation}
The scattering length $a$ can be arbitrary large, while the other
length scales $r_0$, $r_1$, $\ldots$ are assumed to all have
natural sizes, $r_n\sim 1/\Lambda^{(2n+1)}$. Substituting Eq.
(\ref{cot}) in Eq. (\ref{Ap}) we can expand the scattering
amplitude about $p=0$:
\begin{eqnarray}\label{Ar}
{\cal A}(p)&=&-\frac{4\pi
a}{m}[1-iap+(ar_0/2-a^2)p^2\\
&+&O(p^3/\Lambda^3)].\nonumber
\end{eqnarray}
The radius of convergence of this expansion is about $1/|a|$, and
hence in the case $|a|>>1/\Lambda$, which is of relevance to
realistic $NN$ scattering, the expansion shown in Eq. (\ref{cot})
breaks down for momenta far below $\Lambda$. In this case, one
needs to expand the scattering amplitude in powers of $p/\Lambda$
while retaining $ap$ to all orders:
\begin{eqnarray}
A(p)&=&-\frac{4\pi}{m}\frac{1}{a^{-1}+ip}[1+\frac{r_0/2}{a^{-1}+ip}p^2\nonumber\\
&+&\frac{(r_0/2)^2}{(a^{-1}+ip)^2}p^4+\frac{r_1/2}{a^{-1}+ip}p^4+\ldots].\nonumber
\end{eqnarray}
This is the Kaplan-Savage-Wise (KSW) expansion of the effective
theory  \cite{Kaplan2}. It is the expansion that we wish to
reproduce starting with the expression for the $T$-matrix given by
Eq. (\ref{T-}). Using Eq. (\ref{T-}), for the scattering amplitude
$$
A(p)=-\langle{\bf p}| T(z=E_p+i0)|{\bf p}\rangle,
$$
 we can write
\begin{eqnarray}
{\cal
A}(p)=-\frac{|\psi(p/\Lambda)|^2}{C_0^{-1}+\frac{m}{4\pi}ip-M(E_p)}.
\label{A1S0}
\end{eqnarray}
In order to perform the momentum expansion of this amplitude, let
us represent $M(z)$ as follows
\begin{widetext}
\begin{eqnarray}
M(z) =
zm^2\int\frac{d^3k}{(2\pi)^3}\frac{F_1(k/\Lambda)}{(zm-k^2)k^2}
=-zm^2\int\frac{d^3k}{(2\pi)^3}\frac{F_1(k/\Lambda)}{k^4}
+z^2m^3\int\frac{d^3k}{(2\pi)^3}\frac{F_1(k/\Lambda)}{(zm-k^2)k^4}\nonumber\\
=\ldots= -\sum \limits_{n=1}^{\infty}\left((zm)^{n}{\cal
J}_{n}-\frac{m}{4\pi}\sqrt{-zm}\sum
\limits_{i=0}^{n}c^*_{2i}c_{2(n-i)}\left(zm\right)^n\right) =\sum
\limits_{n=1}^{\infty}M_n(z),\label{Mnz}
\end{eqnarray}
\end{widetext}
with
\begin{equation} M_n(z)=-(zm)^{n}{\cal
J}_{n}+\frac{m}{4\pi}\sqrt{-zm}\left(zm\right)^n\sum
\limits_{i=0}^{n}c^*_{2i}c_{2(n-i)},
\end{equation}
where
\begin{eqnarray}
{\cal J}_{n} = m\int\frac{d^3k}{(2\pi)^3}\frac{F_n(k/\Lambda)}{k^{2(n+1)}}\nonumber\\
=m\Lambda^{-2n+1}\int\frac{d^3\tilde{k}}{(2\pi)^3}
\frac{F_n(\tilde{k})}{\tilde{k}^{2(n+1)}}.
\end{eqnarray}
Here $ \tilde{k}=k/\Lambda$,
\begin{eqnarray}
F_n(k/\Lambda)=|\psi(k/\Lambda)|^2-\sum \limits_{i=0}^{n-1}\sum
\limits_{j=0}^{i}c^*_{2j}c_{2(i-j)}k^{2i}.
\end{eqnarray}
 By using Eq. (\ref{Mnz}), we can represent the scattering
amplitude in the form
\begin{eqnarray}
{\cal A}(p)=-\frac{\sum
\limits_{n=0}^{\infty}C_{2n}p^{2n}}{1+\frac{m}{4\pi}ip\sum
\limits_{n=0}^{\infty}C_{2n}p^{2n}},\label{mu=o}
\end{eqnarray}
where $C_{2n}$ are determined by the recurrence relation
\begin{eqnarray}
C_{2n}=C_0\left(\sum \limits_{i=0}^{n}c^*_{2i}c_{2(n-i)}-\sum
\limits_{j=1}^{n}C_{2(n-j)}{\cal J}_{j}\right).\nonumber
\end{eqnarray}
Equation (\ref{mu=o}) allows one to expand the scattering
amplitude as
\begin{eqnarray}
{\cal A}(p)=-\frac{C_0}{1+\frac{m}{4\pi}C_0}-\frac{C_2p^2}
{\left(1+\frac{m}{4\pi}C_0\right)^2}\nonumber\\
+\frac{\frac{m}{4\pi}(C_2)^2ip^5}{\left(1+\frac{m}{4\pi}C_0\right)^3}-
\frac{C_4p^4}{\left(1+\frac{m}{4\pi}C_0\right)^2}+\ldots\nonumber
\end{eqnarray}
Comparing with Eq. (\ref{Ar}), this expression relate the
couplings $C_{2n}$ to the low energy scattering data $a$, $r_n$:
\begin{eqnarray}
C_{0}=\frac{4\pi a}{m},\quad C_2=\frac{2\pi}{m} a^2r_0,\nonumber\\
C_{4}=\frac{4\pi
a^3}{m}\left(\frac{r_0^2}{4}+\frac{r_1}{2a}\right),\ldots
\label{Car}
\end{eqnarray}
By using the expression
\begin{eqnarray}
\frac{1}{\sum
\limits_{n=0}^{\infty}C_{2n}(\mu)p^{2n}}+\frac{m}{4\pi}\mu=\frac{1}{\sum
\limits_{n=0}^{\infty}C_{2n}p^{2n}},\label{expression}
\end{eqnarray}
which relates couplings $C_{2n}(\mu)$ to $C_{2n}(0)= C_{2n}$ as
\begin{eqnarray}
C_{2n}(\mu)=\frac{C_{2n}+\frac{m}{4\pi}\mu\sum
\limits_{i=0}^{n-1}C_{2i}(\mu)C_{2(n-i)}}{1-\frac{m}{4\pi}\mu
C_0},\nonumber
\end{eqnarray}
Eq. (\ref{mu=o}) can be rewritten in the form
\begin{eqnarray}
{\cal A}(p)=-\frac{\sum
\limits_{n=0}^{\infty}C_{2n}(\mu)p^{2n}}{1+\frac{m}{4\pi}(ip+\mu)\sum
\limits_{n=0}^{\infty}C_{2n}(\mu)p^{2n}}.\label{mu}
\end{eqnarray}
This equation is exactly what Kaplan, Savage, and Wise have
derived  \cite{Kaplan2} in the pionless theory by using the PDS
substraction scheme  with the substraction point $p^2=-\mu^2$, and
Eq. (\ref{mu}) yields the following expressions for $C_0(\mu)$ and
$C_2(\mu)$:
\begin{eqnarray}
C_0(\mu)&=&\left(\frac{1}{C_0}-\frac{m}{4\pi}\mu\right)^{-1}=\frac{4\pi/m}{1/a-\mu},\nonumber\\
C_2(\mu)&=&C_2\left(\frac{C_0(\mu)}{C_0}\right)^2=\frac{4\pi}{m}\frac{r_0/2}{(1/a-\mu)^2}.\nonumber
\end{eqnarray}
As for the couplings $C_{2n}(\mu)$ for $n>2$, from Eq.
(\ref{expression}) it follows that they satisfy the equation
$$\mu\frac{d}{d\mu}C_{2n}(\mu)=\frac{m\mu}{4\pi}\sum
\limits_{m=0}^{n}C_{2m}(\mu)C_{2(n-m)}(\mu),
$$
i.e., satisfy the complete, coupled RG equation.

We have demonstrated that the expression for the $2N$ $T$-matrix
shown in Eq. (\ref{T-}) can be used for calculating the scattering
amplitude up to some order in $Q/\Lambda$. As we have seen, Eq.
(\ref{T-}) reproduces the same expansion of the scattering
amplitude that in the standard pionless EFT is derived by summing
the bubble diagrams and performing regularization and
renormalization. On the other hand, in general one cannot restrict
oneself only to the scattering amplitudes. In the next section we
will show that the formalism of the GQD provides a new way to
formulate the effective theory of nuclear forces as a perfectly
consistent theory that allows one to construct not only scattering
amplitudes but also the off-shell $2N$ $T$-matrix, and the Green
and evolution operators.

\section{The effective theory of nuclear forces.}
As follows from Eq. (\ref{evo}), in order to describe the low
energy dynamics we have to obtain the $T$-matrix for any $z$
relevant for the low energy theory. Equation (\ref{difer}) allows
one to obtain the operator $T(z)$ for any $z$ provided that some
boundary condition on this equation is specified. The boundary
condition (\ref{T(z)to}) means that the most of the contribution
to the operator $T(z)$ in the limit $|z|\to\infty$ comes from the
processes associated with the fundamental interaction in the
system under study  that are described by the operator $B(z)$. On
the other hand, for the low energy theory to be consistent, the
operator $B(z)$ must be determined in terms of low energy degrees
of freedom. This means that, despite the boundary condition
(\ref{T(z)to}) formally implies that $z$ must be let to infinity,
really one has to restrict oneself to a "high" energy region which
 is much above the low energy
scale but is well below the scale of the underlying high energy
physics. We will use ${\cal D}$ to denote this "high" energy
region. The fact that the energy region ${\cal D}$ lies much above
the scale of the low energy dynamics  implies that at such
"infinite" $z$ the most of the contribution to the operator $T(z)$
comes from processes that can be thought of as a "fundamental"
interaction in the low energy theory, and, as a result, the
interaction operator $B(z)$ is  close enough to the true $2N$
$T$-matrix. Thus, in order to take into account the fact that any
theory has a range of validity, instead of the boundary condition
(\ref{T(z)to}), we have to use the following boundary condition on
Eq. (\ref{difer}):
\begin{eqnarray}
 \langle \psi_2|T(z)|\psi_1\rangle &=& \langle \psi_2|B(z)|\psi_1\rangle\label{T(z=s)}\\
 &+&O\left\{h(z)\right\},\quad z\in{\cal D}.\nonumber
\end{eqnarray}
For Eq. (\ref{difer}) with the boundary condition (\ref{T(z=s)})
has a unique solution the interaction operator must be close
enough inside domain ${\cal D}$ to the true $2N$ $T$-matrix. This
means that it must satisfy the equation
\begin{eqnarray}
\frac{d \langle \psi_2|B(z)|\psi_1\rangle}{dz} &=& - \sum
\limits_{n} \frac{\langle \psi_2|B(z)|n\rangle\langle n |B(z)|\psi_1\rangle}{(z-E_n)^2}\nonumber\\
 &+&O\left\{|z|^{-1}h(z)\right\},\quad z\in{\cal D}. \label{diferB2}
\end{eqnarray}

 In the low energy theory of nuclear forces all
processes that in ChPT are described by irreducible diagrams
involving only two external nucleons can be considered as such
"fundamental" processes. Here irreducible diagrams are $2N$
irreducible: Any intermediate state contains at least one pion or
isobar. It is natural to expect that the relative contribution of
reducible $2N$ diagrams tends to zero as $z$ increases, and, in a
region that lies much above the scale of low energy nucleon
dynamics, the main contribution comes from the "fundamental"
processes that are described by the $2N$ irreducible diagrams,
i.e., the operator $B(z)$ becomes very close to the true $2N$
$T$-matrix. On the other hand, because of the separation of scales
provided by QCD the  above "high" energy region of the low energy
theory lies still much below the scale of the underlying physics.
In other words, the interaction operator and amplitudes $\langle
\psi_2|B(z)|\psi_1\rangle$ that constitute the interaction
operator and generate low energy nucleon dynamics really are low
energy (in the scale of the underlying theory) amplitudes that in
ChPT are described by the irreducible $2N$ diagrams. In principle
they can be obtained within the underlying high energy theory in
some low energy limit. It is important that the matrix elements of
the $NN$ interaction operator are just the amplitudes that are
described by these diagrams, and hence there is no a need to make
\textit{a priory} assumption that what we have to extract from the
analysis of the diagrams in ChPT is the $NN$ potential. The
amplitudes that are described by these diagrams can be directly
used as  building blocks for constructing the low energy theory.
Thus the GQD allows one to build a bridge between QCD and low
energy nucleon dynamics. It is hoped that in the future it will be
possible to obtain the amplitudes $\langle
\psi_2|B(z)|\psi_1\rangle$ in terms of QCD with such accuracy that
the corresponding operator $B(z)$ will determine a unique low
energy $2N$ $T$-matrix. For this the operator $B(z)$ obtained in
this way must be  close enough to the true $T$-matrix inside the
domain ${\cal D}$. There is no reason to believe that this
operator must necessarily generate the Hamiltonian low energy
dynamics. In fact, as it follows from Eq. (\ref{B(z)-H_I}), for
this $\langle {\bf p}_2|B(z)|{\bf p}_1\rangle$ must have a
negligible dependence on $z$ inside  the domain ${\cal D}$.
However, as we will show below, such a behavior of $\langle {\bf
p}_2|B(z)|{\bf p}_1\rangle$ is at variance with the symmetries of
QCD. On the other hand, QCD must produce the low energy $2N$
$T$-matrix that satisfies the generalized dynamical equation. From
this in turn it follows that QCD must give rise to the $NN$
interaction operator $B(z)$ satisfying the condition
(\ref{diferB2}).

In order to demonstrate how Eq. (\ref{diferB2}) and the analysis
of diagrams in ChPT can be used for constructing the interaction
operator, let us turn to the pionless theory. As has been noted,
from this analysis  it follows that the $2N$ $T$-matrix in the
${}^1S_0$ channel must be of the form (\ref{pp}). Obviously the
interaction operator $B(z)$ that reproduces this $T$-matrix inside
the domain ${\cal D}$ must have the same structure
\begin{equation}
\langle {\bf p_2}|B(z)|{\bf p_1}\rangle = \sum
\limits_{n,m=0}^\infty p_2^{2n}p_1^{2m}b_{nm}(z). \label{bnm}
\end{equation}
Correspondingly from the requirement that the operator $B(z)$ have
such a structure it follows that this operator should be of the
form
\begin{eqnarray}
\langle{\bf p}_2| B(z)|{\bf
p}_1\rangle=\psi^*(p_2/\Lambda)\psi(p_1/\Lambda)f(z),\label{Bf}
\end{eqnarray}
where  $\psi(p/\Lambda)$  is given by Eq. (\ref{psi=psi1}), and
the function $f(z)$ is a solution of the equation
\begin{eqnarray}
\frac{d f(z)}{dz} &=& - f^2(z)J(z)\label{fNdif}\\
&+&O\left\{|z|^{-1}h(z)\right\},\quad z\in{\cal D},\nonumber
\end{eqnarray}
with
\begin{eqnarray}
J(z)=\int\frac{d^3k}{(2\pi)^3}
\frac{|\psi(k/\Lambda)|^2}{(z-E_k)^2}.\nonumber
\end{eqnarray}
Solving Eq. (\ref{fNdif}) yields
\begin{eqnarray}
f(z)=-{\cal M}^{-1}(z)-C_0^{-1}{\cal M}^{-2}(z) \label{calM}\\
 -C_0^{-2}{\cal M}^{-3}(z)\ldots+O\left\{h(z)\right\},
\quad z\in{\cal D},\nonumber
\end{eqnarray}
where ${\cal M}(z)=M_0(z)+M(z)$, with
$M_0(z)=\frac{m}{4\pi}\sqrt{-zm}$. Obviously the number of the
terms we have to keep in this expansion depends on the value of
the parameter $\beta$, i.e., on the accuracy with which the
operator $B(z)$ of the form (\ref{bnm}) reproduces the relevant
solution of Eq. (\ref{difer}) in the limit $ |z|\to\infty$. Since
the function ${\cal M}(z)$ is completely determined by the form
factor $\psi(p/\Lambda)$, all solutions of Eq. (\ref{fNdif}) for
given form factor have the same first term in the expansion
(\ref{calM}), and only the second term distinguishes the different
solutions (they are fixed by the free parameter $C_0$). Thus in
order that Eq. (\ref{difer}) with the interaction operator of the
form (\ref{bnm}) has a unique solution, we have to keep the first
two terms in the expansion of the function $f(z)$ shown in Eq.
(\ref{calM}). This means that the function $h(z)$ in Eq.
(\ref{fNdif}) is determined by the equation
\begin{eqnarray}
h(z)=O\left\{|C_0^{-2}{\cal
M}^{-3}(z)|\right\}=O\left\{|C_0^{-2}M_0^{-3}(z)|\right\}.\label{power}
%\nonumber
\end{eqnarray}
From the above it follows that the operator describing the $NN$
interaction in the ${}^1S_0$ channel can be written as
\begin{eqnarray}
\langle{\bf p}_2| B(z)|{\bf p}_1\rangle=
-\psi^*(p_2/\Lambda)\psi(p_1/\Lambda)\label{BzM}\\
\times\left({\cal M}^{-1}(z)+C_0^{-1}{\cal
M}^{-2}(z)\right).\nonumber
\end{eqnarray}
 Correspondingly, for
the interaction operator $H_{int}^{(s)}(\tau)$, we have
\begin{eqnarray}
\langle{\bf p}_2| H_{int}^{(s)}(\tau)|{\bf
p}_1\rangle&=&\psi^*(p_2/\Lambda)\psi(p_1/\Lambda)\nonumber\\
&\times& \left(f_1(\tau)+C_0^{-1}f_2(\tau)\right),\nonumber
\end{eqnarray}
where
\begin{eqnarray}
f_n(\tau)=\frac{i}{2\pi}\int\limits_{-\infty}^{\infty}dx\exp(-iz\tau)
{\cal M}^{-n}(z).\nonumber
\end{eqnarray}

As we have shown, the knowledge of the form factor in Eq.
(\ref{Bf}) completely determines the form of the $NN$ interaction
operator which in turn determines the dynamics of the system.
Equation (\ref{difer}) with this interaction operator has a unique
solution, and this solution is the $T$-matrix shown in Eq.
(\ref{T-}). However, this implies that we know the details of the
dynamics at high energies, because the relevant domain of
definition of the function $\psi(p/\Lambda)$ spreads to such
energies. At the same time, the basic premise of effective
theories is that low energy dynamics can be described in terms of
a few parameters without any knowledge of the details of high
energy interactions. As we have seen in Sec.IV, the functions
$\psi(p/\Lambda)$ and $M(z)$ that enter the expression for the
operator $B(z)$ can be expanded as is shown in Eqs.
(\ref{psi=psi1}) and  (\ref{Mnz}), and in order to calculate these
functions and hence $\langle{\bf p}_2|B(z)|{\bf p}_1\rangle$ with
accuracy up to the order $(Q/\Lambda)^{2{\cal N}}$ it is
sufficient to know the first $N$ terms in these expansions. In
other words, for determining the interaction operator with the
accuracy up to the order $(Q/\Lambda)^{2{\cal N}}$, one need only
to know the properties of the form factor that are parameterized
by the constants $c_{2n}$, ... $c_{2{\cal N}}$ and ${\cal J}_1$,
..., ${\cal J}_{\cal N}$. Let us now show
 that, starting with the interaction operator
(\ref{BzM}), where only the first $N$ terms in the expansion of
$\psi(p/\Lambda)$ and the same in the expansion of $M(z)$ are
known, one can calculate the $T$-matrix with the accuracy up to
the order $(Q/\Lambda)^{2{\cal N}}$ for all relevant $z$. For
calculating with this accuracy, instead of the interaction
operator (\ref{BzM}), one can use the following effective
interaction operator
\begin{widetext}
\begin{eqnarray}
\langle{\bf p}_2| B^{({\cal N})}_{eff}(z)|{\bf
p}_1\rangle=-\left(1+c_2^*p_2^2+\ldots+c_{2{\cal N}}^*p_2^{2{\cal
N}}+\varphi_{\cal N}^*(p_2/\Lambda)\right)
\left(1+c_2p_1^2+\ldots+c_{2{\cal N}}p_1^{2{\cal N}}+\varphi_{\cal N}(p_1/\Lambda)\right)\nonumber\\
\times\left({\cal M}^{-1}_{\cal N}(z)+C_0^{-1}{\cal M}^{-2}_{\cal
N}(z)\right),\label{Beffn}
\end{eqnarray}
\end{widetext}
with
\begin{eqnarray}
&&{\cal M}_{\cal N}(z)=\sum\limits_{n=0}^{{\cal N}}M_n(z)\nonumber\\
&=&\frac{m}{4\pi}\sqrt{-zm}\sum\limits_{n=0}^{{\cal
N}}(zm)^n\sum\limits_{i=0}^{n}c_{2i}^*c_{2(n-i)}
-\sum\limits_{n=1}^{{\cal N}}(zm)^{n}{\cal J}_{n},\nonumber
\end{eqnarray}
where  $\varphi_{\cal N}(p/\Lambda)$ is an arbitrary function that
satisfies the condition
\begin{eqnarray}
\varphi_{\cal
N}(p/\Lambda)=O\left\{\left(p/\Lambda\right)^{2({\cal
N}+1)}\right\}.\label{phi-order}
\end{eqnarray}
The form of this effective interaction operator manifests
uncertainty  in the details of the $NN$ interaction at high
energies. In contrast with the operator $B(z)$ which unequally
determines the relevant solution of Eq. (\ref{difer}), the
effective interaction operator $B_{eff}^{({\cal N})}(z)$
determines the set of the solutions that coincide inside the
domain ${\cal D}$ with the true $2N$ $T$-matrix with the relative
accuracy up to the order $(Q/\Lambda)^{2({\cal N}+1)}$
\begin{eqnarray}
\langle{\bf p}_2| T(z)|{\bf p}_1\rangle&=&\langle{\bf p}_2|
B^{({\cal N})}_{eff}(z)|{\bf
p}_1\rangle\label{inconBeff}\\
\times\left(1+O(Q/\Lambda)^{2({\cal
N}+1)}\right)&+&O\left\{|C_0^{-2}M_0^{-3}(z)|\right\},\quad
z\in{\cal D}.\nonumber
\end{eqnarray}
This implies that the operator $B_{eff}^{({\cal N})}(z)$ must
satisfy Eq. (\ref{difer}) with the same accuracy
\begin{widetext}
\begin{eqnarray}
\frac{d\langle{\bf p}_2| B_{eff}^{({\cal N})}(z)|{\bf
p}_1\rangle}{dz}%\\
=-\int\frac{d^3k}{(2\pi)^3}\frac{\langle{\bf p}_2| B_{eff}^{({\cal
N})}(z)|{\bf k}\rangle\langle{\bf k}| B_{eff}^{({\cal N})}(z)|{\bf
p}_1\rangle}{(z-E_k)^2}\nonumber\\
\times\left(1+O(Q/\Lambda)^{2({\cal
N}+1)}\right)+O\left\{|C_0^{-2}|z|^{-1}M_0^{-3}(z)|\right\},\quad
z\in{\cal D}.\label{difBeff}
\end{eqnarray}
By taking into account Eqs. (35-37), one can easily verify that
this condition is really satisfied, provided the function
$\varphi_{\cal N}(p/\Lambda)$ is such that
\begin{eqnarray}
m\int\frac{d^3k}{(2\pi)^3}\frac{ \left|1+c_2k^2+\ldots+c_{2{\cal
N}}k^{2{\cal N}}+\varphi_{{\cal N}}(k/\Lambda)\right|^2-
\sum\limits_{i=0}^{n-1}\sum\limits_{j=0}^{i}c_{2j}^*c_{2(i-j)}k^{2i}}{k^{2(n+1)}}={\cal
J}_{n},\label{calJn}
\end{eqnarray}
\end{widetext}
where $1\leq n \leq {\cal N}$. Thus the requirement that inside
the domain ${\cal D}$ the effective interaction operator is close
enough to the true $2N$ $T$-matrix determines some properties of
the function $\varphi_{{\cal N}}(p/\Lambda)$ that is the unknown
part of the form factor $\psi(p/\Lambda)$. The functions
$\varphi_{{\cal N}}(p/\Lambda)$ mainly characterize the form
factor at high energies. For this reason, for describing low
energy nucleon dynamics with the accuracy up to the order
$O(Q/\Lambda)^{2({\cal N}+1)}$ one need not to know the function
$\varphi_{\cal N}(p/\Lambda)$ exactly. This is because the details
of the high energy physics must not affect on the low energy
dynamics. It is sufficient to know the integration properties of
this function shown in Eq. (\ref{calJn}). Thus the constants
${\cal J}_1$, ..., ${\cal J}_{\cal N}$ in the effective
interaction operator $B_{eff}^{({\cal N})}(z)$ parameterize the
effects of high energy physics on low energy nucleon dynamics.

By definition, the effective interaction operator $B_{eff}^{({\cal
N})}(z)$ is not so close inside the domain ${\cal D}$ to the true
$2N$ $T$-matrix to determine a unique solution of Eq.
(\ref{difer}). However, it determines a set of the solutions
$T^{({\cal N})}(z)$ of Eq. (\ref{difer}) that coincide with the
true $2N$ $T$-matrix, with the relative accuracy up to the order
$O(Q/\Lambda)^{2({\cal N}+1)}$ (we will denote this set
$\Omega_{\cal N}$). Each of these solutions corresponds to some
definite function $\varphi_{\cal N}(p/\Lambda)$ satisfying the
conditions (\ref{phi-order}) and (\ref{calJn}) and can be
represented as
\begin{eqnarray}
\langle{\bf p}_2| T^{({\cal N})}(z)|{\bf p}_1\rangle&=&t_{\cal
N}(z)\left(\sum\limits_{n=0}^{\cal N}
c_{2n}^*p_2^{2n}+\varphi_{\cal N}^*(p_2/\Lambda)\right)\nonumber\\
&\times&\left(\sum\limits_{n=0}^{\cal N}
c_{2n}p_1^{2n}+\varphi_{\cal N}(p_1/\Lambda)\right),\nonumber
\end{eqnarray}
where $t_{\cal N}(z)$ is the solution of the equation
\begin{eqnarray}
\frac{dt_{\cal N}(z)}{dz}=-t_{\cal N}^2(z){\cal J}(z)=t_{\cal
N}^2(z)\frac{d{\cal M}(z)}{dz},\label{dM}
\end{eqnarray}%\end{widetext}
with the boundary condition
\begin{eqnarray}
t_{\cal N}(z)=-\left({\cal M}_{\cal N}^{-1}(z)+C_0^{-1}{\cal
M}^{-2}_{\cal
N}(z)\right)&&\label{66}\\
\times\left(1+O\left\{\left(Q/\Lambda\right)^{2({\cal
N}+1)}\right\}\right)+O\left\{|C_0^{-2}M_0^{-3}(z)|\right\},\nonumber\\\quad
z\in&{\cal D}.& \nonumber
\end{eqnarray}
The general solution of Eq. (\ref{dM}) is
\begin{eqnarray}
t_{\cal N}(z)= -\frac{1}{C^{-1}-{\cal M}(z)},\label{C}
\end{eqnarray}
where $C$ is some arbitrary constant. Equation (\ref{C}) may be
rewritten in the form
\begin{eqnarray}
t_{\cal N}(z)= -\frac{1}{C^{-1}-{\cal M}_{\cal N}(z)-\tilde{\cal
M}_{\cal N}(z)},\nonumber
\end{eqnarray}
where $\tilde{\cal M}_{\cal N}(z)\equiv\sum\limits_{n={\cal
N}}^\infty M_n(z)$ is the part of ${\cal M}(z)$ that depends on
the choice of the function $\varphi_{\cal N}(p/\Lambda)$. From
this equation it follows that
\begin{eqnarray}
t_{\cal N}(z)= -\left({\cal M}_{\cal N}^{-1}(z)+C^{-1}{\cal
M}_{\cal
N}^{-2}(z)\right)\nonumber\\
\times\left(1+O\{(Q/\Lambda)^{2({\cal N}+1)}\}\right)+O\left\{|C^{-2}M_0^{-3}(z)|\right\},%\nonumber\\
\quad z\in{\cal D}.\nonumber
\end{eqnarray}
Here we have used the fact that $\tilde{\cal M}_{\cal N}(z)$ is of
order $O\left\{\left(Q/\Lambda\right)^{2({\cal N}+1)}\right\}$. On
the other hand the relevant solution of Eq. (\ref{dM}) must
satisfy the boundary condition (\ref{66}). From this it follows
that the parameter $C$ in Eq.  (\ref{C}) must be equal to $C_0$.
Thus the solution of Eq. (\ref{difer}) with the boundary condition
given by Eqs. (\ref{Beffn}) and (\ref{inconBeff}) where
$\varphi_{\cal N}(p/\Lambda)$ is specified in some way is
\begin{widetext}
\begin{eqnarray}
\langle{\bf p}_2| T^{({\cal N})}(z)|{\bf p}_1\rangle&=&
\frac{\left(\sum\limits_{n=0}^{\cal N}
c_{2n}^*p_2^{2n}+\varphi_{\cal N}^*(p_2/\Lambda)\right)
\left(\sum\limits_{n=0}^{\cal N} c_{2n}p_1^{2n}+\varphi_{\cal
N}(p_1/\Lambda)\right)}{C_0^{-1}-{\cal M}_{\cal N}(z)-\tilde{\cal
M}_{\cal N}(z)}
\label{69}
\\&=&\frac{\left(\sum\limits_{n=0}^{\cal
N} c_{2n}^*p_2^{2n}+\varphi_{\cal N}^*(p_2/\Lambda)\right)
\left(\sum\limits_{n=0}^{\cal N} c_{2n}p_1^{2n}+\varphi_{\cal
N}(p_1/\Lambda)\right)}{C_0^{-1}-{\cal M}_{\cal
N}(z)}\left(1+O\left\{\left(Q/\Lambda\right)^{2({\cal
N}+1)}\right\}\right).\nonumber
\end{eqnarray}
\end{widetext}
However, as has been noted, the function $\varphi_{\cal
N}(p/\Lambda)$ that enters  the effective interaction operator is
not known exactly. At a given order, only its integral properties
shown in Eq. (\ref{calJn}) are known. For this reason  Eq.
(\ref{69}) actually represents the set $\Omega_{\cal N}$ of the
solutions of Eq. (\ref{difer}) which coincide each with other with
accuracy up to the order $O\left\{\left(Q/\Lambda\right)^{2({\cal
N}+1)}\right\}$, and the true $2N$ $T$-matrix belongs to this set.
This is because the "operator" $B_{eff}^{({\cal N})}(z)$ actually
is a set of the interaction operators $B(z)$ which coincide with
the true $2N$ interaction operator with the same accuracy. This
uncertainty  in specifying the boundary condition on the
generalized dynamical equation is only difference of the effective
theory from a "fundamental" one whose predictions are assumed to
be exact. It is important that in both of these cases we deal with
the same well defined equation of motion which can be used not
only for analytic, but also for numerical calculations.

As an example, let us consider the effective pionless theory at
next-to-leading order $({\cal N}=1)$. In this case, for the
effective interaction operator, we have
\begin{eqnarray}
&&\langle{\bf p}_2| B_{eff}^{(1)}(z)|{\bf
p}_1\rangle=-\left(1+c_2^*p_2^2+\varphi_1^*(p_2/\Lambda)\right)\label{B^1}\\
&\times&\left(1+c_2p_1^2+\varphi_1(p_1/\Lambda)\right) \left({\cal
M}_1^{-1}(z)+C_0^{-1}{\cal M}_1^{-2}(z)\right),\nonumber
\end{eqnarray}
where
$${\cal M}_1(z)=\frac{m}{4\pi}\sqrt{-zm}-zm{\cal J}_1-\frac{m}{2\pi}(-zm)^{3/2}{\text Re}{}c_2,$$
and $\varphi_1(p/\Lambda)$ is some function satisfying the
condition shown in Eq. (\ref{calJn}). The solution of Eq.
(\ref{difer}) with this effective interaction operator yields
\begin{eqnarray}
\langle{\bf p}_2| T^{(1)}(z)|{\bf
p}_1\rangle=\left(1+c_2^*p_2^2+\varphi_1^*(p_2/\Lambda)\right)\label{T^1}\\
\times\frac{\left(1+c_2p_1^2+\varphi_1(p_1/\Lambda)\right)}
{C_0^{-1}-{\cal M}_1(z)}%\nonumber\\
\left(1+O\left\{(Q/\Lambda)^4\right\}\right).\nonumber
\end{eqnarray}
This equation represents the set of the solutions of the
generalized dynamical equation that coincide with the true $2N$
$T$-matrix with the relative accuracy up to the order
$(Q/\Lambda)^4$. At this order one may use any of these solutions.
Correspondingly the operator $B_{eff}^{(1)}(z)$ with any function
$\varphi_1(p/\Lambda)$ satisfying the conditions (\ref{phi-order})
and  (\ref{calJn}) may be used as the interaction operator $B(z)$
that determines a unique solution of Eq. (\ref{difer}). For
example, one may chose this function as follows
$\varphi_1(p/\Lambda)=c_2p^2\exp(-p^2/\Lambda_1^2)$, with
$\Lambda_1=2\pi\sqrt{\pi}{\cal J}_1(m{\text Re}{}c_2)^{-1}$.

It should be noted that the expression (\ref{69}) which represents
the form of the effective $2N$ $T$-matrix in the ${}^1S_0$ channel
of the pionless theory can also be  derived from Eq. (\ref{T-}) by
keeping only the first ${\cal N}$ terms in both the expansions of
$\psi(p/\Lambda)$ and $M(z)$ that enter in this equation. This is
because in the case of the pionless theory the generalized
dynamical equation can be solved exactly, and as we have shown,
the unique solution consistent with symmetries of QCD is the
$T$-matrix (\ref{T-}). However, this is not the case when pions
must be included as explicit degrees of freedom. In the theory
with pions, one cannot solve the problem exactly, and hence the
way of constructing the $2N$ $T$-matrix presented in Sec.III is
inapplicable. Above we have presented the way of constructing the
effective theory that prescribes to start with the analysis of the
theory at "high" energies belonging to the domain ${\cal D}$. At
these energies the theory has the most simple structure. For
example, in the case where the dynamics in the theory is
Hamiltonian, $\langle \psi_2|T(z)|\psi_1\rangle$ have a negligible
dependence on $z$ inside the domain ${\cal D}$. This means that
the interaction generating the dynamics can be assumed to be
instantaneous and hence the low energy dynamics can be described
by an interaction Hamiltonian that is just the value of the
$T$-matrix in this domain. If we deal with several different
interactions, then the interaction Hamiltonian is the sum of the
corresponding interaction Hamiltonians. This is because processes
that involve several different interactions cannot occur
instantaneously, and hence give a negligible contribution to the
$T$-matrix inside the domain ${\cal D}$. In the general case,
however, the operator $T(z)$ have significantly depend on $z$ even
inside this domain. As it follows from Eqs.  (\ref{T-}) and
(\ref{69}), this takes place in the case of the contact $NN$
interaction.

As an illustration of the advantages of the above formulation of
the effective theory, let us consider the dynamics of the $2N$
system in the presence of an external potential. For reason of
physical transparency, we will consider nucleons as spinless
particles whose interaction is described by the interaction
operator  (\ref{BzM}), i.e., is the same as in the ${}^1S_0$
channel of the pionless theory. The interaction operator
generating the dynamics of such a system should be of the form
\begin{eqnarray}
\langle{\bf P}_2;{\bf p}_2| B^{pot}(z)|{\bf P}_1;{\bf
p}_1\rangle&=&\langle{\bf P}_2;{\bf p}_2| B(z)|{\bf P}_1;{\bf
p}_1\rangle\nonumber\\&+&\langle{\bf P}_2;{\bf p}_2|V|{\bf
P}_1;{\bf p}_1\rangle,\label{Bpot}
\end{eqnarray}
where the vector $|{\bf P};{\bf p}\rangle$ describes the state of
the $2N$ system with the momentum of the center of mass of the
system ${\bf P}$ and the relative momentum of nucleons ${\bf p}$.
The operator $B(z)$ that describes the interaction of nucleons
with themselves and, in the center of mass frame, is shown in Eq.
 (\ref{BzM}). The second term on the right hand side of Eq. (\ref{Bpot})
describes the instantaneous interaction with the external
potential. We will assume that this potential is local, i.e., has
the form
\begin{eqnarray}
\langle{\bf P}_2;{\bf p}_2|V|{\bf P}_1;{\bf
p}_1\rangle%\nonumber\\
=V(q)(2\pi)^3\delta^{(3)}({\bf p}_2-{\bf p}_1-{\bf q})\nonumber
\end{eqnarray}
with ${\bf q}={\bf P}_2-{\bf P}_1$, and is week enough to be
considered in the Born approximation.

Let the details of the $NN$ interaction are known only up to
next-to-leading order. In this case we have to solve the problem
in the spirit of the effective theory starting with the effective
interaction operator of the form
\begin{eqnarray}
\langle{\bf P}_2;{\bf p}_2| B^{pot}_{eff}(z)|{\bf P}_1;{\bf
p}_1\rangle=\langle{\bf P}_2;{\bf p}_2| B^{(1)}_{eff}(z)|{\bf
P}_1;{\bf p}_1\rangle\nonumber\\+V(q)\delta^{(3)}({\bf p}_2-{\bf
p}_1-{\bf q}),\label{Bpot=B^1}
\end{eqnarray}
where $B_{eff}^{(1)}(z)$ is given by Eq. (\ref{B^1}). This
operator determines a set of the solutions of Eq. (\ref{difer})
that coincide with the true $T$-matrix describing the dynamics of
the system with accuracy up to the order $(Q/\Lambda)^4$. It can
be checked that in the Born approximation these solutions are
given by the equation
\begin{eqnarray}
T^{pot}(z)=T^{(1)}(z)\nonumber\\
+\left(1+T^{(1)}(z)G_0(z)\right)V\left(1+G_0(z)T^{(1)}(z)\right),\label{Tpot}
\end{eqnarray}
where $T^{(1)}(z)$ is one of the solutions of  Eq. (\ref{difer})
shown in Eq. (\ref{T^1}). By using Eq. (\ref{Tpot}), one can
construct, with a given accuracy, the scattering amplitude and the
evolution operator. For example, for the amplitude of  $2N$
scattering in the presence of the external potential, we can write
\begin{eqnarray}
A({\bf P}_2,{\bf p}_2;{\bf P}_1,{\bf p}_1)=-\langle{\bf P}_2;{\bf
p}_2|\left(1+T^{(1)}(z)G_0(z)\right)\label{A^1}\\\times
V\left(1+G_0(z)T^{(1)}(z)\right)|{\bf P}_1;{\bf
p}_1\rangle,\nonumber
\end{eqnarray}
where
$z=\frac{P_1^2}{4m}+\frac{p_1^2}{m}+i0=\frac{P_2^2}{4m}+\frac{p_2^2}{m}+i0$,
and ${\bf p}_1\neq{\bf p}_2$. By using Eq. (\ref{T^1}), for the
scattering amplitude $A({\bf P}_2,{\bf p}_2;{\bf P}_1,{\bf p}_1)$
we get
\begin{widetext}
\begin{eqnarray}
A({\bf P}_2,{\bf p}_2;{\bf P}_1,{\bf p}_1)=A_{00}({\bf P}_2,{\bf
p}_2;{\bf P}_1,{\bf p}_1) +A_{01}({\bf P}_2,{\bf p}_2;{\bf
P}_1,{\bf p}_1)&+&A_{10}({\bf P}_2,{\bf p}_2;{\bf P}_1,{\bf
p}_1)+A_{11}({\bf P}_2,{\bf p}_2;{\bf P}_1,{\bf p}_1),\label{Ampl}
\end{eqnarray}
where
\begin{eqnarray}
A_{00}({\bf P}_2,{\bf p}_2;{\bf P}_1,{\bf
p}_1)=-V(q)(2\pi)^3\delta^{(3)}({\bf p}_2-{\bf p}_1-{\bf
q}),\nonumber
\end{eqnarray}
\begin{eqnarray}
A_{01}({\bf P}_2,{\bf p}_2;{\bf P}_1,{\bf
p}_1)=-\frac{V(q)\langle{\bf p} |T^{(1)}(z-E_{P_1})|{\bf
p}_1\rangle}{z_1-E_{p}-E_{P_1}}\left(1+O(Q/\Lambda)^4\right)\nonumber\\
=-\frac{V(q)(1+c_2^*p^2+c_2p_1^2)\left(1+O(Q/\Lambda)^4\right)}{(E_{p_1}-E_{p})\left(
C_0^{-1}+\frac{m}{4\pi}ip_1 +{\cal
J}_1p_1^2+\frac{m}{2\pi}ip_1^3{\text Re}c_2\right)},\qquad {\bf
p}={\bf p}_2-{\bf q},\nonumber
\end{eqnarray}
\begin{eqnarray}
A_{10}({\bf P}_2,{\bf p}_2;{\bf P}_1,{\bf
p}_1)=-\frac{V(q)\langle{\bf p}_2|T^{(1)}(z-E_{P_2})|{\bf
p'}\rangle}{z-E_{p'}-E_{P_2}}\left(1+O(Q/\Lambda)^4\right)
\nonumber\\
=-\frac{V(q)(1+c_2^*p_2^2+c_2p'^2)\left(1+O(Q/\Lambda)^4\right)}{(E_{p_2}-E_{p'})\left(
C_0^{-1}+\frac{m}{4\pi}ip_2 +{\cal
J}_1p_2^2+\frac{m}{2\pi}ip_2^3{\text
Re}{}c_2\right)},\quad\qquad{\bf p'}={\bf p}_1+{\bf q},\nonumber
\end{eqnarray}
\begin{eqnarray}
A_{11}({\bf P}_2,{\bf p}_2;{\bf P}_1,{\bf p
}_1)=-V(q)\int\frac{d^3p}{(2\pi)^3}\frac{\langle{\bf
p}_2|T^{(1)}(z-E_{P_2})|{\bf p'}\rangle\langle{\bf
p}|T^{(1)}(z-E_{P_1})|{\bf
p}_1\rangle}{(z-E_{ p'}-E_{P_2}+i0)(z-E_{p}-E_{P_1}+i0)}\left(1+O(Q/\Lambda)^4\right)\nonumber\\
=-\frac{V(q)\left(1+c_2^*p_2^2+c_2p_1^2\right)}{\left(
C_0^{-1}+\frac{m}{4\pi}ip_2 +{\cal
J}_1p_2^2+\frac{m}{2\pi}ip_2^3{\text
Re}{}c_2\right)\left(C_0^{-1}+\frac{m}{4\pi}ip_1 +{\cal
J}_1p_1^2+ \frac{m}{2\pi}ip_1^3{\text Re}{}c_2\right)}\nonumber\\
\times[\int\frac{d^3p}{(2\pi)^3}\frac{1}
{(E_{p_2}-E_{{p'}}+i0)(E_{p_1}-E_{p}+i0)}+\int\frac{d^3p}{(2\pi)^3}\frac{(E_p-E_{p'}+E_{p_2})
(c_2^*p^2+c_2^*p'^2)} {E_p(E_{p_2}-E_{{
p'}}+i0)(E_{p_1}-E_{p}+i0)}\nonumber\\-E_{p_1}\int\frac{d^3p}{(2\pi)^3}\frac{(c_2^*p^2+c_2^*p'^2)}
{E_p^2(E_{p_1}-E_{p}+i0)}+m{\cal
J}_1]\left(1+O(Q/\Lambda)^4\right),\qquad {\bf p'}={\bf p}+{\bf
q}. \label{A11}
\end{eqnarray}
\end{widetext}
Here we have used the fact that the function
$\varphi_1(p/\Lambda)$ satisfies the condition (\ref{calJn}). Thus
the generalized dynamical equation with the effective $NN$
interaction operator (\ref{Beffn}) allows one to calculate, with
the desired accuracy, not only $2N$ data but also processes in the
presence of an external field. As we have seen, in this way
regularization and renormalization are not required, and hence one
need not to introduce new parameters: The solution depends only on
the constants $C_0$, $c_2$, and ${\cal J}_1$ that are contained in
the effective interaction operator. As we have noted, the coupling
$C_0$ is fixed by the scattering length $C_0=\frac{4\pi}{m}a$.
However, knowing the $2N$ scattering data is not sufficient to
obtain the parameter $c_2$ and ${\cal J}_1$. Only the coupling
$C_2=C_0(c_2+c^*_2-C_0{\cal J}_1)=C_0\frac{ar_0}{2}$ is fixed by
the scattering data at this order,  but not the couplings $c_2$
and ${\cal J}_1$ separately. At the same time, as it follows from
Eq. (\ref{A11}), these parameters manifest themselves in  $2N$
scattering  in the presence of an external potential. This is
because the $2N$ $T$-matrices enter half-of-the-energy shell in
the scattering amplitude (\ref{A^1}) describing such processes. In
other words, such processes can be used for obtaining the above
couplings.

\section{The probabilistic frame of quantum mechanics and
the character of low energy nucleon dynamics.}

In an EFT renormalization allows one to describe low energy
physics using an effective Lagrangian that contains only a few
degrees of freedom, because after renormalization that consists in
absorbing infinities in a redefinition of constants in the
Lagrangian the remaining integrals are effectively cut off at
internal momenta lager than $Q$. This might lead us to the
conclusion that the definition of any effective theory must
necessarily include regularization and renormalization. However,
as we have shown, by using the example of the pionless theory,
regularization and renormalization are needed when we describe the
theory in terms of an effective Lagrangian. But, if we consider
the problem from the more general point of view provided by the
GQD and do not restrict ourselves to the assumption that the
effective action is instantaneous, we see that the effective
theory of nuclear forces manifests itself as a perfectly
consistent theory free from the UV divergences. As we have shown,
in the pionless theory the $2N$ $T$-matrix in the ${}^1S_0$
channel is given by Eq. (\ref{T-}). This $T$-matrix satisfies the
generalized dynamical equation, i.e., the solution of this
equation does not require regularization and renormalization. In
other words, in the pionless theory the generalized dynamical
equation allows one to separate the low energy physics  from the
underlying high energy physics without renormalization. Let us now
show that this directly follows from the first principles of
quantum mechanics.

The fact that because of the separation of scales, at  low
energies nucleons emerge as the only effective degrees of freedom
means that the temporal evolution of the system can be described
in terms of the $2N$ Hilbert subspace ${\cal H}_{2N}$ and the
evolution  operator defined on this space
\begin{eqnarray}
U_{2N}(t,0)=P_{2N}U_{S}(t,0)P_{2N}, \label{proek}
\end{eqnarray}
where $P_{2N}$ being the projection operator on the subspace
${\cal H}_{2N}$. Here we use the Schr{\"o}dinger picture. Since in
this case the $2N$ system is considered to be closed, the
evolution operator $U_{2N}(t,0)$ should be unitary
\begin{eqnarray}
U^+_{2N}(t,0)U_{2N}(t,0)& = 1. \label{unitcond}
\end{eqnarray}
This is one of the main requirements that quantum mechanics
imposes on such a theory. It expresses the fact that, if at
initial time the system was in the state $|\psi\rangle\in{\cal
H}_{2N}$, then in a measurement at time $t$ the system will be
necessarily found in one of the $2N$ states. Here, of course, it
is assumed that ${\cal H}_{2N}$ describes $2N$ states at low
energies. The above means that the sum of the probabilities to
find the system in all the possible $2N$ states must be equal to
unite
\begin{equation}
\int \frac{d^3k}{(2\pi)^3} |\langle{\bf
k}|U_{2N}(t,0)|{\bf\psi}\rangle|^2=1,\label{limit}
\end{equation}
with $|\psi\rangle$ being a normalized vector belonging to ${\cal
H}_{2N}$. Here we consider nucleons as spinless particles, and $
|\langle{\bf
k}|U_{2N}(t_2,t_1)|{\bf\psi}\rangle|^2\frac{d^3k}{(2\pi)^3}$ is
the probability of finding the quantum system in the $2N$ state
with the relative momentum ${\bf p}$ in the volume of the momentum
space $k_i\leq p_i\leq k_i+dk_i$ $(i=1,2,3)$. Finally, for Eq.
(\ref{limit}) to be valid for any normalized vector
$|\psi\rangle\in{\cal H}_{2N}$, the operator $U_S(t,0)$ must be
unitary.

Another basic principle of quantum mechanics is the principle of
the superposition of the probability amplitudes from which it
follows that the evolution operator is defined by Eq. (\ref{U-G})
which in this case reads
\begin{eqnarray}
\langle{\bf p}_2|U_{2N}(t,0)|{\bf p}_1\rangle=
\frac{i}{2\pi}\int\limits_{-\infty}^\infty dx \exp(-izt)
G_{2N}(z),\label{U2N}
\end{eqnarray}
with
\begin{eqnarray}
G_{2N}(z)=G'_0(z)+G'_0(z)T_{2N}(z)G'_0(z),\label{G2N}
\end{eqnarray}
where
\begin{eqnarray}
G'_0(z)=P_{2N}G_0(z)P_{2N}\label{G'}
\end{eqnarray}
 and $T_{2N}(z)$ is defined by the
equation
\begin{eqnarray}
\langle{\bf p}_2|T_{2N}(z)|{\bf p}_1\rangle &=
&i\int\limits_0^\infty d(t_2-t_1)
\exp[iz(t_2-t_1)]\nonumber\\&\times& \langle{\bf
p}_2|\widetilde{T}_{2N}(t_2-t_1)|{\bf
p}_1\rangle,\label{T2N(z)=T2N(t}
\end{eqnarray}
Thus, in the case of the reduced $2N$ system, we deal with the
same equations as in describing the complete system. The only
difference is that the amplitudes $\langle{\bf
p}_2|\widetilde{T}_{2N}(t_2-t_1)|{\bf p}_1\rangle$ describes the
contributions to the evolution operator from the processes in
which the interaction in the $2N$ system (not in the complete
system) begins at time $t_1$ and ends at time $t_2$. By using Eq.
(\ref{U2N}), the unitarity condition (\ref{unitcond}) may be
rewritten as
\begin{eqnarray}
\frac{1}{(2\pi)^2}\int\limits_{-\infty}^{\infty}dx_1\int\limits_{-\infty}^{\infty}dx_2\exp[it(z_2-z_1)]
\nonumber\\
\times\langle\psi_2|G^+_{2N}(z_2)G_{2N}(z_1)|\psi_1\rangle
&=&\langle\psi_2|\psi_1\rangle.\label{cond}
\end{eqnarray}
For Eq. (\ref{cond}) to be valid for any time $t$, the $2N$
$T$-matrix must satisfy Eq. (\ref{difer}). Note that in Ref.
\cite{R.Kh.:1999} the generalized dynamical equation has been
derived just in this way. At the same time, the above does not
mean that $T_{2N}(z)$ satisfies the LS equation.  In other words,
from the existence of the separation of scales which results in
the unitarity of the $2N$ evolution operator $U_{2N}(t,0)$ and the
superposition principle it follows that the $2N$ $T$-matrix must
necessarily satisfy the generalized dynamical equation but not the
LS equation. Thus the dynamics of the $2N$ system may be
non-Hamiltonian.

Let us now show that the above is true even if the dynamics in the
underlying theory is governed by the Schr{\"o}dinger equation. As
we have noted, in the case where the dynamics of a quantum system
is Hamiltonian the evolution operator $U_S(t,0)$ can be
represented in the form (\ref{U-G}) with the Green operator given
by Eq. (\ref{G}). From this it follows that
\begin{eqnarray}
& &U_{2N}(t,0)=P_{2N}U_{S}(t,0)P_{2N}\nonumber\\
& = &\frac{i}{2\pi}\int_{-\infty}^{\infty}dx\exp(-izt)P_{2N}(z-H)^{-1}P_{2N}\nonumber\\
& = &\frac{i}{2\pi}\int_{-\infty}^{\infty}dx\exp(-izt)G_{2N}(z).
 \label{pek}
\end{eqnarray}

It can be shown, that the Green operator
$G_{2N}=P_{2N}(z-H)^{-1}P_{2N}$ may be represented in the form
\begin{eqnarray}%\label{P2N}
G_{2N}(z)=P_{2N}(z-H)^{-1}P_{2N}\label{proe}\\
=G'_0(z)+G'_0(z)T_{2N}(z)G'_0(z),\nonumber
\end{eqnarray}
where $G'_0(z)$ is defined by Eq. (\ref{G'}).  The above means
that, if the underlying dynamics is governed by the
Schr{\"o}dinger equation, then the $2N$ evolution operator
$U_{2N}(t,0)$ can be represented in the form
\begin{eqnarray}
& &U_{2N}(t,0)=\frac{i}{2\pi}\int_{-\infty}^{\infty}dx\exp(-izt)\nonumber\\
&\times&\left(G'_0(z)+G'_0(z)T_{2N}(z)G'_0(z)\right).\label{re}
\end{eqnarray}
From this it follows that $T_{2N}(z)$ must satisfy Eq.
(\ref{difer}), because the operator (\ref{re}) is assumed to be
unitary. At the same time, the above does not mean that the
amplitudes $\langle{\bf k_2}|T(z)|{\bf k_1}\rangle$ fall off
rapidly enough above the scale of the effective theory for the
dynamics in the theory to be governed by the Schr{\"o}dinger
equation. In fact, the Hamiltonian character of the underlying
dynamics implies only that the $2N$ $T$-matrix must fall off
rapidly above the scale of the underlying physics.
 Thus, the
dynamics in the effective low energy theory may be non-Hamiltonian
even if  the underlying high energy dynamics is governed by the
Schr{\"o}dinger equation.

An important lesson to be learned from the above analysis is  that
low energy nucleon dynamics  can be described by the generalized
dynamical equation reduced to the $2N$ Hilbert space without
regularization and renormalization. The divergence problems may
arise only in describing the dynamics of the effective theory by
using the Schr{\"o}dinger (LS) equation, because such a
description is based on the additional assumption that the
effective $NN$ interaction is instantaneous and, as a consequence,
the generalized dynamical equation is equivalent to the
Schr{\"o}dinger equation. Note, in this connection, that from the
physical point of view the fact that the chiral potentials lead to
UV divergences means that the LS equation with such potentials
does not provide a well separation of the low energy physics from
the underlying high energy physics. In other words, the amplitudes
$\langle {\bf k}_2|T(z)|{\bf k_1}\rangle$ does not fall off
rapidly enough above the scale of the effective theory for
integrals in the LS equation to be effectively cut off at internal
momenta larger than $Q$. In this case, in order to integrate out
the high energy degrees of freedom, one need to use regularization
and renormalization. In contrast, the generalized dynamical
equation should separate the low energy physics that is described
by the effective theory of nuclear forces from the underlying high
energy physics provided that at very low energies nucleons may
emerge as the only explicit degrees of freedom. In fact, the
letter means that the matrix elements of the operator
$U_{2N}(t,0)$ must fall off rapidly enough at momenta larger than
$Q$. This is because the probability of finding, for a measurement
at time $t$, the system in the $2N$ state with momenta larger than
$Q$, if at time $t=0$ the system was in the low energy state
$|\psi\rangle$, must be negligible. If this is not the case, one
cannot also ignore other states with such momenta in which the
high energy degrees of freedom manifest themselves, and hence the
unitarity condition (\ref{limit}) breaks down. As it follows from
Eqs. (\ref{difer}), (\ref{U-G}) and (\ref{G-T}), the fact that
 $\langle {\bf k}_2|U(t,t_0)|{\bf
k_1}\rangle$ fall off rapidly at momenta larger than $Q$ in turn
means that the integrals in the generalized dynamical equation
effectively cut off at such momenta, and hence, in this case the
high energy degrees of freedom can be integrated out without
renormalization.

\section{A new look at the Weinberg program}

Our formulation of the effective theory of nuclear forces may be
regarded as a new way of realizing the Weinberg program for
physics of the two-nucleon systems. The implementation of this
program has three stages. Firstly, one must consider the case
where the $NN$ system is not subject to any external probes and
employ ChPT to generate a nonrelativistic
particle-number-conserving Hamiltonian for the nuclear system. At
the second stage this Hamiltonian should be employed for
constructing the full $2N$ $T$-matrix via the LS (the
Schr{\"o}dinger) equation. The third step of the Weinberg program
is to employ the $2N$ $T$-matrix obtained in this way in the case
where $2N$ system is subject some external probe: a pion, a
photon, or some weakly-interacting particle. Provided this probe
carries momentum of order $m_\pi$ its interaction with the $2N$
system can be represented as a sum of irreducible diagrams which
forms a kernel $K_{probe}$ for the process of interest. The full
amplitude for this process is then found by multiplying this
kernel $K_{probe}$ by the factors describing the interaction of
the $NN$ pair in the initial and final states
\begin{eqnarray}
A=(1+TG_0)K_{probe}(1+G_0T).
\end{eqnarray}

The Weinberg proposal was based on the assumption that the only
equation that can govern low energy nucleon dynamics is the LS
(the Schr{\"o}dinger) equation, and hence what one has to derive
from the analysis of diagrams in ChPT is an $NN$ potential.
However, there is no reason to consider that low energy nucleon
dynamics is necessarily governed by the Schr{\"o}dinger equation.
In principle this dynamics may be governed by the generalized
dynamical equation with a nonlocal-in-time interaction operator
when this equation is not equivalent to the Schr{\"o}dinger
equation. In fact, as has been shown in Ref. \cite{R.Kh.:1999},
only the generalized dynamical equation must be satisfied in any
case, not the Schr{\"o}dinger equation. In the light of this fact
the Weinberg program can be considered from a new point of view:
Instead of the Schr{\"o}dinger (LS) equation, one should use the
generalized dynamical equation. In this case one need not to use
$a$ $priory$ assumption that from the analysis of diagrams for the
$2N$ $T$-matrix it follows that the an effective $NN$ interaction
is instantaneous.  Such a modification of the Weinberg program is
natural and does not change its character. Indeed, the fact that
the analysis of the diagrams in ChPT leads to singular chiral
potentials in the case of which the LS equation makes no sense
without renormalization means that really this equation is not
sufficient for constructing the $2N$ $T$-matrix. In addition one
need to specify a renormalization scheme to make these predictions
finite. But the LS equation plus a renormalization scheme is
something more general than this equation itself, and, after
renormalization we deal with the dynamics that is governed by
another equation which, as has been shown in Ref.
\cite{R.Kh.:1999}, may be only the generalized dynamical equation.
In order to illustrate this point let us consider the effective
theory at leading order of the Weinberg power counting. The $NN$
effective potential that has been derived from the Weinberg
analysis of diagrams in ChPT is \cite{EFT3}
\begin{eqnarray}
V({\bf p}_2,{\bf
p}_1)=&-&\left(\frac{g_A^2}{2f_\pi^2}\right)\frac{\bf{q}
\cdot{\bf{\sigma_1}}{\bf{q}} \cdot{\bf \sigma}_2}
{q^2+m_\pi^2}{\bf\tau_1}\cdot{\bf{\tau}_2}\nonumber\\&+&C_S+C_T{\bf{\sigma_1}}
\cdot{\bf{\sigma}_2},\nonumber
\end{eqnarray}
with ${\bf q}\equiv{\bf p}_2-{\bf p}_1$. The coupling $g_A$ is the
axial coupling constant, $m_\pi$ is the pion mass, $f_\pi$ is the
pion decay constant, and ${\bf \sigma}({\bf \tau})$ are the Pauli
matrices acting in spin (isospin) space. At very low energies the
one-pion-exchange part of the $NN$ interaction may be included
into the contact term, and the $NN$ potential takes the form
\begin{equation}
\langle{\bf p}_2|V|{\bf p}_1\rangle=C,\label{sing}
\end{equation}
where $C_0=C_S-3C_T+g_A^2/f_\pi^2$.  Obviously, the potential
(\ref{sing}) is singular, and the LS equation with this potential
makes no sense without regularization and renormalization. Let us
perform regularization by using a momentum cut-off. In this case
the singular potential (\ref{sing}) is replaced by the regularized
one
%\begin{eqnarray}
$\langle{\bf p}_2|V_\Lambda|{\bf
p_1}\rangle=f^*(p_2/\Lambda)C_0(\Lambda)f(p_1/\Lambda),$% \label{cut}
%\end{eqnarray}
where the form factor $f(p/\Lambda)$ satisfies $f(0)=1$ and falls
off rapidly for $p/\Lambda>1$. The solution of the LS equation
with this potential is
\begin{eqnarray}
&&\langle{\bf p_2}|T_{\Lambda}(z)|{\bf
p_1}\rangle=f^*(p_2/\Lambda)f(p_1/\Lambda)\nonumber\\
&\times&\left(C_0^{-1}(\Lambda)-\int\frac{d^3k}{(2\pi)^3}
\frac{|f(k/\Lambda)|^2}{z-E_k}\right)^{-1}.\nonumber
\end{eqnarray}
Defining the renormalized value $C_R$ of $C_0$ as the value of the
$T$-matrix $\langle{\bf p_2}|T_{\Lambda}(z)|{\bf p_1}\rangle$ at
$z=\frac{p_1^2}{m}=\frac{p_2^2}{m}=0$
\begin{eqnarray}
C_R^{-1}=C_0^{-1}(\Lambda)+\int\frac{d^3k}{(2\pi)^3}
\frac{|f(k/\Lambda)|^2}{E_k},\label{-1}
\end{eqnarray}
we may rewrite this equation as
\begin{eqnarray}
\langle{\bf p_2}|T_{\Lambda}(z)|{\bf
p_1}\rangle&=&\frac{f^*(p_2/\Lambda)f(p_1/\Lambda)}{C_R^{-1}-z\int\frac{d^3k}{(2\pi)^3}
\frac{|f(k/\Lambda)|^2}{(z-E_k)E_k}}, \nonumber
\end{eqnarray}
where the renormalized value is fixed by the scattering length
$a=mC_R/4\pi$. Thus, after renormalization the integral in the
expression for the leading order $2N$ $T$-matrix is effectively
cut off, and hence at this stage the regularization may be removed
by letting $\Lambda\to\infty$, and we get
\begin{eqnarray}
\langle{\bf p_2}|T(z)|{\bf p_1}\rangle=
\left(C_R^{-1}(\Lambda)-\frac{m^{3/2}}{4\pi}\sqrt{-z}\right)^{-1}.
\label{eq2}
\end{eqnarray}
Correspondingly, for the half-of-the-energy shell $T$ matrix, we
have
\begin{eqnarray}
\langle{\bf p'}|T(E_p+i0)|{\bf p}\rangle=
\left(C_R^{-1}+\frac{im|{\bf p}|}{4\pi}\right)^{-1}. \nonumber
\end{eqnarray}
This is just what has been obtained by Weinberg in Ref.
\cite{EFT3}. Obviously the $T$-matrix (\ref{eq2}) is not a
solution of the LS equation. At the same time, as we have seen,
this $T$-matrix is the well defined solution of the generalized
dynamical equation with the interaction operator (\ref{hint})
describing a nonlocal in time interaction. The dependence of this
interaction operator on $z$ expresses the fact that the
instantaneous interaction which is described by the contact
potential (\ref{sing}) does not contribute to the $2N$ $T$-matrix
separately. In fact, as it follows from Eq. (\ref{-1})  after
removing regularization, this potential becomes equal to zero. A
nonzero contribution to this $T$-matrix comes only from the sum of
diagrams involving infinite number of such contact interactions.
In other words renormalization spreads effective contact $NN$
interaction in time.

Thus even if we start with the LS equation and try to extract an
$NN$ potential from the analysis of diagrams in ChPT, despite it
is singular and makes no sense, after renormalization we come to
low energy nucleon dynamics that is governed by the generalized
dynamical equation with a nonlocal-in-time interaction operator.
This means that in order to realize the Weinberg program in a
consistent way one has to use this equation instead of the LS
equation. As we have shown in the previous section, in contrast
with the LS equation, the generalized dynamical equation separates
the low energy physics from the high energy physics without
renormalization. This manifests itself in the fact that, being
reduced to describing the $2N$ system, this equation is free from
UV divergences. In Sec.V we have demonstrated the advantages of
this way of realizing the Weinberg program by using the example of
the pionless theory. Firstly, by using the analysis of diagrams in
ChPT, we have obtained the effective $NN$ interaction operator. In
the ${}^1S_0$ channel it is of the form (\ref{Beffn}). By solving
the generalized dynamical equation with this interaction operator
we have constructed the ${}^1S_0$ channel $2N$ $T$-matrix shown in
Eq.  (\ref{69}).  It is important that in this way we have
obtained not only the $2N$ scattering amplitude but also the
off-shell $T$-matrix. Equation (\ref{69}) expresses the constrains
that the symmetries of QCD place on the off-shell behavior of the
$2N$ $T$-matrix. At a given order there are only a few free
parameters in Eq. (\ref{69}) that can be derived from low energy
experiment. Once they are obtained in any way, the off-shell $2N$
$T$-matrix that in this case is completely determined by Eq.
(\ref{69}) can be used in Eq. (\ref{A^1}) for describing the
interaction of the $2N$ system with an external probe. In Sec.V we
have demonstrated this fact by using the example of the $2N$
scattering in the presence of an external potential. As we have
seen, with the $2N$ $T$-matrix (\ref{69}) in hand the full
amplitudes of the processes involving the interaction with
external probes can be calculated without regularization and
renormalization.

The key point of the above way of the realization of the Weinberg
program is the derivation of the $NN$ interaction operator from
the analysis of the time-ordered diagrams in the time-ordered
diagrams in the energy region ${\cal D}$. This region is
sufficiently above the scale of the low energy physics for the
processes that are described by the $2N$ $T$-matrix at $z\in{\cal
D}$ may be considered as a "fundamental" interaction in this
theory. At the same time, this energy region is much below the
scale of the underlying high energy physics, and hence the
effective $NN$ interaction operator is really determined by the
low energy in the scale of QCD behavior of the time-ordered
diagram for $2N$ $T$-matrix. Moreover, one need not to know this
behavior exactly. It is sufficient to know the structure of this
$T$-matrix, i.e., its dependence on momenta of nucleons, that is
predicted by ChPT. For example, in the pionless theory the
${}^1S_0$ channel $2N$ $T$-matrix has the structure shown in Eq.
(\ref{pp}). Correspondingly the structure of the interaction
operator is given by Eq.  (\ref{bnm}). On the other hand, this
interaction must be close enough in the energy region ${\cal D}$
to the dynamical equation, i.e., must satisfy Eq. (\ref{diferB2}).
This requirement yields the expression for the effective
interaction operator shown in Eq.  (\ref{Bf}). This interaction
operator can then be used for constructing the $2N$ $T$-matrix. At
the same time, in the pionless theory the requirement that the
$2N$ $T$-matrix having the structure (\ref{pp}) satisfy the
generalized dynamical equation directly yields the expression for
the $2N$ $T$-matrix shown in Eq.  (\ref{T-}), and this expression
can be used for organizing the calculations of the effective $2N$
$T$-matrix. This is because the $2N$ $T$-matrix has the same
simple structure at all relevant $z$ as in the "high" energy
region ${\cal D}$. However, this is not the case in the theory
with pions or (and) when one must take into account the Coulomb
interaction between protons. In such theories the structure of the
$2N$ $T$-matrix is much simpler in the "high" energy region than
at low energies where the nonperturbative character of nucleon
dynamics manifests itself. In this case only in the region ${\cal
D}$ the $2N$ $T$-matrix, with the accuracy needed for constructing
the interaction operator, can be represented as a sum of
contributions from the contact, pion-exchange, and Coulomb
interactions. The fact that starting from this high energy
representation one can construct the $NN$ interaction operator and
then use them for obtaining the $2N$ $T$-matrix has been
demonstrated in Ref. \cite{archive} by using the example of the
$2N$ dynamics at leading order of the Weinberg power counting. As
we have seen, the same situation takes place in the case when one
consider the $2N$ dynamics in presence of an external potential.
In this case interaction operator generating the dynamics in the
system is given by Eq.  (\ref{Bpot}), and its form manifests the
fact that the "high" energy region ${\cal D}$ the main
contribution to the $2N$ $T$-matrix can be represented as a sum of
the contributions from the interaction of nucleons with themselves
and the external potential. By using the interaction operator
(\ref{Bpot=B^1}) one can describe the dynamics of this system.

\section{The off-shell behavior of the $2N$ $T$-matrix and
the three-nucleon problem}

As we have shown, the formulation of the effective theory of
nuclear forces presented in Sec. V allows one to  construct not
only the $2N$ scattering amplitude (in this case we reproduce all
results of the standard EFT approach), but also the off-shell
$T$-matrix, and the evolution and Green operators. This is very
important because the $S$-matrix is not everything. For example,
at finite temperature there is no $S$-matrix because particles
cannot get out to infinite distances from a collision without
bumping into things. This means that the off-shell structure of
the $2N$ $T$-matrix should influence on in-medium observables. In
Ref. \cite{Fuchs} it has been shown that the off-shell behavior of
the in-medium nucleon-nucleon $T$-matrix and hence the off-shell
properties of nuclear forces have substantial effects on the
transition amplitudes and cross sections at large nuclear matter
densities. These properties are crucial for solving the
many-nucleon problem. For example, the $2N$ amplitudes enter
off-the-energy-shell in the $3N$ equations.

The realistic $NN$ potentials that describe $2N$ scattering data
to high precision can not guarantee that a similar precision will
be achieved in the description of larger nuclear systems. In fact,
the simplest observable in the $3N$ system, the binding energy of
the triton, is under predicted by the realistic $NN$ potentials
which are so successful in describing the $2N$ observables. The
energy deficit ranges from 0.5 to 0.9 MeV and depends on the
off-shell and short-range parameterization of the $2N$ force
\cite{Kievsky}. In order to resolve this problem one has to take
into account three-nucleon force (3NF) contributions to the $3N$
binding energy. The common way of solving the $3N$ bound state
problem is to use in the Schr{\"o}dinger equation phenomenological
$NN$ potentials and then to introduce a 3NF to provide
supplementary binding. However, from the point of view of the
three-nucleon problem, it is not sufficient to generate a
phenomenological $NN$ potential that perfectly reproduces the $2N$
scattering amplitudes. One must also generate a $NN$ potential by
using theoretical insight as much as possible in order to
constrain the off-shell properties of the $2N$ $T$-matrix. If this
is not the case, a $NN$ potential which fits precisely the $2N$
phase shifts but produces the erroneous off-shell behavior of the
$T$-matrix would not provide reliable results for the $3N$ system,
nor can be used to test for the presence of $3N$ forces. It is
important, in this context, that the requirement that the $2N$
$T$-matrix satisfy the generalized dynamical equation and has the
form consistent with the symmetries of QCD places a constrain on
the off-shell behavior of this $T$-matrix. As we have shown, in
the pionless theory the $2N$ $T$-matrix in the ${}^1S_0$ channel
should be of the form (\ref{T-}). The expression shown in Eq.
(\ref{T-}) contains the infinite set of parameters $c_{2n}$ that
in principle could be obtained in terms of QCD. At the same time,
as we have seen, the generalized dynamical equation allows one to
organize calculations of the two-nucleon $T$-matrix in the spirit
of the effective theory by parameterizing the effects of the
underlying physics in a few parameters that can be derived from a
low energy experiment. In this way we have obtained the expression
(\ref{69}) that determines $2N$ $T$-matrix in the ${}^1S_0$
channel with the accuracy up to the order $(Q/\Lambda)^{2({\cal
N}+1)}$. For example, at next-to-leading order the $T$-matrix is
given by Eq. (\ref{T^1}) containing only three free parameters
${\cal J}_1$, ${\text Re}{}c_2$ and ${\text Im}{}c_2$ (the
parameter $C_0$ is fixed by the scattering length at leading
order). One of them is fixed by the $2N$ phase shifts analysis
while other two can be derived from the in-medium scattering data
or from the low energy data corresponding to the processes of the
interaction of the two-nucleon system with an external probe. In
other words, the requirement that the $2N$ $T$-matrix be
consistent with basic principles of quantum mechanics and the
symmetries of QCD removes the off-shell ambiguities. This may
provide a better understanding of the many-nucleon problem.

The models for the $3N$F that are usually used for solving the
problem with the triton understanding are based on two-pion
exchange with intermediate $\Delta$-isobar excitation. However,
these $3N$F models cannot explain the $A_y$ puzzle. In Ref.
\cite{Canton2} a three-nucleon force generated by the exchange of
one pion in the presence of a $2N$ correlation \cite{Canton} has
been suggested as a possible candidate to explain the $A_y$
puzzle. This $3N$F contribution is fixed by the $2N$ $T$-matrix
describing the underlying 2N interaction while the pion is "in
flight". The expression for this force derived in Ref.
\cite{Canton} contains the off-shell $2N$ $T$-matrix, more
precisely its subtracted part
\begin{equation}
  \tilde t_{12}({\bf p}_2,{\bf p}_1,z)=t_{12}({\bf p}_2,{\bf p}_1,z)-v_{12}({\bf p}_2,{\bf
  p}_1),\label{t_12}
\end{equation}
where ${\bf p}_1$ and ${\bf p}_2$ are Jacobi momenta of nucleons 1
and 2, and the potential - like term $v_{12}({\bf p}_2,{\bf p}_1)$
contains only OPE/OBE - type diagrams. The subtraction in Eq.
(\ref{t_12}) is needed to take into account a cancellation effect
which has been observed in Refs. \cite{Yang,Yang2}. This
cancellation involves meson retardation effects of the iterated
Born term, and the irreducible diagrams generated by sub-summing
all time ordering diagrams describing the combined exchange of two
mesons amongst the three nucleons. In principle there are no free
parameters to adjust, and the $3N$ force is completely determined
by the $2N$ $T$-matrix. However, as has been shown in Ref.
\cite{Canton2}, in order to explain the $A_y$ puzzle, instead of
the subtracted $T$-matrix shown in Eq. (\ref{t_12}), one has to
use the amplitude defined according to the prescription
\begin{eqnarray}
  \tilde t_{12}({\bf p}_2,{\bf p}_1,z)=c(z)t_{12}({\bf p}_2,{\bf p}_1,z)-v_{12}({\bf p}_2,{\bf
  p}_1),\nonumber
\end{eqnarray}
with the effective parameter $c(z)$, which represents an overall
correction factor for the far-off-the-energy-shell $2N$
$T$-matrix. Ideally, this parameter should be one for the $2N$
potential to provide a reliable extrapolation of the $2N$
$T$-matrix down to $z\approx -160$ MeV. However, as has been shown
in Ref. \cite{Canton2}, none of the existing $2N$ $T$-matrices can
guarantee the off-shell behavior that is needed for the
explanation of the $A_y$ puzzle with the parameter $c(z)$ set to
one. For example, in order to reproduce the $nd$ experimental data
with the Bonn B potential, the factor $c(z)$ must be set to 0.73
for the energy 3 MeV.

The lesson one must learn from the above is that the off-shell
behavior of the $2N$ $T$-matrix may play a crucial role in
explaining the $A_y$ puzzle, and the existing realistic potentials
do not provide the off-shell behavior that is needed for the
correct reproduction for $A_y$ with the $3N$F suggested in Ref.
\cite{Canton}. Note in this connection that, being fitted to the
$2N$ scattering data, the existing realistic potentials lead to
ambiguities in the off-shell behavior of the $2N$ $T$-matrix. In
order to remove the off-shell ambiguities one has to find the way
of constructing the $2N$ $T$-matrix as an inevitable consequence
of the basic principles of quantum mechanics and the symmetries of
QCD. The remarkable feature of the EFT approach is that it implies
to build the theory of nuclear forces as a consequence of these
principles. However, the standard EFT of nuclear forces does not
predict the off-shell behavior of the $2N$ $T$-matrix. The cause
of this is that in this theory the Schr{\"o}dinger (LS) equation
makes no sense without regularization and renormalization. On the
other hand, there is no reason to restrict ourselves to the
assumption that low energy nucleon dynamics is governed by the
Schr{\"o}dinger equation: Only the generalized dynamical equation
must be satisfied in any case, and, as we have seen, the
requirement that the solutions of this equation that are
consistent with the symmetries of QCD correspond to a
nonlocal-in-time interaction operator when the generalized
dynamical equation cannot be reduced to the Schr{\"o}dinger
equation. At the same time, as has been shown in Ref.
\cite{PRC:2002}, the nonlocality in time of the interaction
results in an anomalous off-shell behavior of the $T$-matrix. In
other words, the formalism of the GQD predicts that in order that
the $2N$ $T$-matrix to be consistent with the symmetries of QCD it
must have the off-shell behavior that cannot take place in the
case of the ordinary potentials. From this point of view, the
"anomalous" off-shell behavior that the $2N$ $T$-matrix must have
for the $3NF$ be able to explain the $A_y$ puzzle may be
considered as a manifestation of the fact that low energy nucleon
dynamics is really non-Hamiltonian.

\section{Summary and Discussion}

We have shown that from the Weinberg analysis of time-ordered
diagrams for the $2N$ $T$-matrix in ChPT it follows that nucleon
dynamics at low energies is governed by the generalized dynamical
equation with a nonlocal-in-time interaction operator. A
remarkable feature of the generalized dynamical equation which
follows straightforwardly from the first principles of quantum
mechanics is that it allows one to construct all physical
amplitudes relevant for the theory under consideration by using
the amplitudes describing processes in which the duration time of
interaction is infinitesimal. It is natural to assume that the
most of contribution to these amplitudes comes from the processes
associated with a fundamental interaction in a quantum system.
This point manifests itself in the boundary condition
(\ref{fund}). If we do not consider a theory that is valid up to
infinitely high energies (infinitesimal times), then the above
infinitesimal of the duration times of interaction should mean
that being much smaller than the scale of the theory these times
may be much larger than the time scale of the underlying high
energy physics. This in turn means that the amplitudes describing
the "fundamental" interaction in the low energy theory can be
computed in terms of the underlying high energy physics. Thus the
generalized dynamical equation allows one to take into account
that every theory with which  we deal is a low energy
approximation to a more fundamental one and provides a bridge
between them.

The generalized dynamical equation can be represented in the form
of the differential equation (\ref{difer}) for the operator
$T(z)$. The boundary condition on the generalized dynamical
equation of this form is shown in Eq. (\ref{T(z)to}), where the
operator $B(z)$ describes the fundamental interaction in this
system. By definition, this operator must be so close to the true
$T$-matrix in the limit $|z|\to\infty$ that the generalized
dynamical equation with the boundary condition (\ref{T(z)to}) have
a unique solution. The above means that really this region of
"infinite" energies with which we have to deal is the domain
${\cal D}$ that lies much above the scale of the low energy
dynamics but much below the scale of the underlying high energy
physics. Correspondingly for describing the low energy dynamics we
have to start with the boundary condition (\ref{T(z=s)}) that
implies that the interaction operator $B(z)$ is so close to the
relevant $T$-matrix inside the domain ${\cal D}$ that Eq.
(\ref{difer}) with this initial condition has a unique solution.
In the theory of nuclear forces this domain is a region of
energies that are high enough for the most of contribution to the
$2N$ $T$-matrix to come from processes that are described by the
irreducible $2N$ diagrams for the $2N$ $T$-matrix (these processes
are associated with a "fundamental" interaction), but not so high
for the heavy degrees of freedom manifest themselves explicitly.
By using the analysis of the time-ordered diagrams for the $2N$
$T$-matrix in ChPT inside the domain ${\cal D}$ where the
structure of the theory is much simpler than in the low energy
region, we can obtain the $NN$ interaction operator which should
be used in the boundary condition (\ref{T(z=s)}) on Eq.
(\ref{difer}).  We have shown that the dynamics which is generated
by the interaction operator obtained in this way is
non-Hamiltonian. This is because, for low energy nucleon dynamics
to be Hamiltonian, the operator $T(z)$ must have a negligible
dependence on $z$ inside the domain ${\cal D}$, while, as we have
shown, this is not the case. In other words, in the
nonrelativistic limit QCD leads through ChPT to low energy nucleon
dynamics that is not governed by the Schr{\"o}dinger equation.
However, this does not mean that the low energy predictions of QCD
are not consistent with quantum mechanics. This means only that in
this case we deal with a nonlocal-in-time interaction when the
generalized dynamical equation cannot be reduced to the
Schr{\"o}dinger equation.

We have shown that from the fact that at extreme low energies
nucleons emerge as the only effective degrees of freedom it
follows that the evolution operator $U_{2N}(t,t_0)$ defined on the
$2N$ subspace ${\cal H}_{2N}$ is unitary and can be represented in
the form  (\ref{T2N(z)=T2N(t}). This in turn means that the $2N$
$T$-matrix that enters  Eq.  (\ref{cond}) must satisfy the
generalized dynamical equation. In other words the dynamics in a
consistent effective theory of nuclear forces must be governed by
the generalized dynamical equation, and the problems with UV
divergences can arise only if we restrict ourselves to the
assumption that the effective $NN$ interaction is instantaneous
and hence the generalized dynamical equation is equivalent to the
Schr{\"o}dinger equation. What we have to do then is to obtain the
interaction operator $B(z)$, i.e., the operator that is so close
to the true solution that the generalized dynamical equation with
the boundary condition (\ref{T(z)to}) has a unique solution.
Ideally, this operator should be derived from QCD. However, for
describing low energy nucleon dynamics with a given accuracy one
need not to know the operator $B(z)$ exactly. It is sufficient to
know the effective interaction operator $B^{({\cal N})}_{eff}(z)$
that determines the solution of Eq. (\ref{difer}) with the same
accuracy. The advantage of using such an interaction operator
whose form manifests the symmetries of QCD is that in this case
the effects of high energy physics on low energy nucleon dynamics
are parameterized by a few constants. The effective interaction
operator $B^{({\cal N})}_{eff}(z)$ determines the set
$\Omega_{\cal N}$ of the solutions of Eq. (\ref{difer}) that
coincide with the true two-nucleon $T$-matrix with the accuracy up
to the order $(Q/\Lambda)^{2{\cal N}}$, i.e., it represents the
set of the interaction operators $B(z)$ that determine the above
solutions. This uncertainty in specifying the initial condition
for the generalized dynamical equation manifests the fact that we
do not know the details of the underlying physics. The effects of
this physics on the low energy dynamics are parameterized by the
constants $c_2$, $\ldots$, $c_{2{\cal N}}$ and ${\cal J}_1$,
$\ldots$, ${\cal J}_{\cal N}$ containing in the operator
$B^{({\cal N})}_{eff}(z)$ that is shown in Eq. (\ref{Beffn}). As
${\cal N}$ increases the set $\Omega_{\cal N}$ of the solutions of
Eq. (\ref{difer}) that are determined by the effective interaction
operator $B^{({\cal N})}_{eff}(z)$ become smaller and smaller.
This means that in this case one approaches to the true solution
closer and closer, because the true $2N$ $T$-matrix enters each of
these sets. Each of the $T$-matrices belonging to the set
$\Omega_{\cal N}$ corresponds to some function $\varphi_{\cal
N}(p/\Lambda)$, which must be such that the effective interaction
operator satisfies Eq. (\ref{diferB2}). This requirement puts the
constraints on the function $\varphi_{\cal N}(p/\Lambda)$: it must
satisfy the condition (\ref{calJn}) where the parameters ${\cal
J}_n$ enter the effective interaction operator through the
function $M_n(z)$. In other words, the effective interaction
operator contains all parameters that appear in the theory at a
given order, and these parameters need not be redefined in the
process of calculations. From this point of view the only
difference of the effective theory from a full one is uncertainty
in specifying the boundary condition on the generalized dynamical
equation, and the boundary condition  (\ref{T(z=s)}) with the
effective interaction operator of the form  (\ref{Beffn}) means
that one may choose any function $\varphi_{\cal N}(p/\Lambda)$
satisfying the conditions (\ref{phi-order}) and (\ref{calJn}).
This is equivalent to the choice of the interaction operator
$B(z)$ that formally determines a unique solution of Eq.
(\ref{difer}). Thus being formulated in terms of the GQD the
effective theory of nuclear forces can be put on the same firm
theoretical grounds as the quantum mechanics of atomic phenomena.
In this case we deal with a well defined equation of motion that
allows one to construct not only the scattering amplitudes but
also the off-shell $T$-matrix, and the evolution and Green
operators.

Our formalism may be regarded as a new way of realizing the
Weinberg program for physics of the two-nucleon systems based on
the use of the generalized dynamical equation instead the LS
equation. In this case at the first stage of the program we derive
from the analysis of the diagrams ChPT an effective interaction
operator. In the ${}^1S_0$ channel of the pionless theory this
operator is of the form (\ref{Beffn}). Solving the generalized
dynamical equation with this interaction operator yields the
expression for the $2N$ $T$-matrix shown in Eq. (\ref{69}). This
$T$-matrix leads to the KSW expansion \cite{Kaplan2} of the
scattering amplitude shown in Eq.  (\ref{expression}). However, in
contrast with standard approach to the EFT of nuclear forces where
this expansion is obtained by summing the bubble diagrams and
performing regularization and renormalization, we reproduce the
same result in a consistent way free from UV divergences. At the
same time, Eq. (\ref{dM}) determines the off-shell $2N$ $T$-matrix
in the ${}^1S_0$ channel up to a few constants that parameterize
the effects of the underlying physics on low energy nucleon
dynamics. In other words, the requirement that low energy nucleon
dynamics be consistent with the symmetries of QCD leads to a
constrain on the off-shell behavior of the $2N$ $T$-matrix, and
one of the lessons that one must learn from the results of our
work is that at very low energies (in the pionless theory) the
${}^1S_0$ off-shell $2N$ $T$-matrix  consistent with the basic
principles of quantum mechanics and the symmetries of QCD must, be
of the form (\ref{69}). With this $T$-matrix in hand
implementation of the third step of the Weinberg program can
proceed without resorting to regularization and renormalization.
For example, formula  (\ref{Ampl}) we have derived allows one to
calculate, in a consistent way free from UV divergences, the $2N$
scattering amplitudes in the presence of an external potential to
subleading order. This potential, for example, may be the Coulomb
potential. Of course, in this case an additional term describing
the contribution from gauge-invariant, four-nucleon, one-photon
operator that arise at next-to-leading order must be included.
However, in our case this term does not play a role of a
regulator, like the counterterms that are used in the standard EFT
approach for calculating such observables as the
electric-quadrupole moment of the deuteron. It should be
emphasized that, if the expression shown in Eq. (\ref{A11}) where
divergent, then this term could not absorb divergences for all
momenta of the nucleons. Moreover, the same off-shell $2N$
$T$-matrix can be used in describing the interaction of the $2N$
system with other probes, in the three-body continuum
calculations, and in solving the microscopic nuclear structure
problems. The advantage of the formulation of the effective theory
of nuclear forces based on the GQD  is that it allows one to
construct the $2N$ interaction operator  and hence the off-shell
$2N$ $T$-matrix as inevitable consequence of the basic principles
of quantum mechanics and the symmetries of QCD. This opens new
possibilities for solving many problems in nuclear physics, and,
in particular, for explaining the $A_y$ puzzle.

In this paper we focused on the pionless theory because in this
simple case the proposed way to formulate the effective theory of
nuclear forces can be investigated in detail, and its ideas can be
verified exactly. At the same time, the approach should be
applicable to the theory with pions. The only problem is that in
this case the definition of a consistent power counting scheme and
the derivation of the effective operator of the $NN$ interaction
from the analysis of the diagrams in ChPT are more complicated.
However, once this operator is constructed the generalized
dynamical equation can be used for performing not only analytic
but also numerical calculations of any observables without
regularization and renormalization.

\end{document}